\documentclass[12pt]{article}
\usepackage[numbers]{natbib}
\usepackage{a4}
\usepackage{citesort}
\usepackage{amsmath}
\usepackage{amssymb}
\usepackage{latexsym}
\usepackage{amssymb}
\usepackage{epsfig}
\usepackage{natbib}
\def \TT{{\mathrm{T}}}
\def \TL{{\mathrm{L}}}

\def \d{{\mathrm{d}}}

\def \R{{\mathbb{R}}}

\def \pd{\partial}

\def \Bsigma{\boldsymbol{\sigma}}
\def \Bbeta{\boldsymbol{\beta}}

\def \BT{\boldsymbol{T}}
\def \BI{\boldsymbol{I}}

\def \Bs{\boldsymbol{s}}

\def \Bb{{\boldsymbol{b}}}

\def \Bu{{\boldsymbol{u}}}

\def \Bv{{\boldsymbol{v}}}
\def \BR{{\boldsymbol{R}}}
\def \BV{{\boldsymbol{V}}}

\def \OO{{\cal{O}}} 
\def \VV{{\cal{V}}}

\def \BJ{{\boldsymbol{J}}}

\def \Bbeta{\boldsymbol{\beta}}
\def \Balpha{\boldsymbol{\alpha}}
\def \rr{{\boldsymbol{r}}}
\def \RR{{\boldsymbol{R}}}
\def \BJ{{\boldsymbol{J}}}
\def \BA{{\boldsymbol{A}}}
\def \BB{{\boldsymbol{B}}}
\def \BE{{\boldsymbol{E}}}

\def \Bu{{\boldsymbol{u}}}
\def \Bv{{\boldsymbol{v}}}

\def \Bp{{\boldsymbol{p}}}

\begin{document}
\title{{\bf 
On the non-uniform motion of dislocations: 
The retarded elastic fields, the retarded dislocation tensor potentials 
and the Li\'enard-Wiechert tensor potentials 
}}
\author{
Markus Lazar~$^\text{a,b,}$\footnote{
{\it E-mail address:} lazar@fkp.tu-darmstadt.de (M.~Lazar).} 
\\ \\
${}^\text{a}$ 
        Heisenberg Research Group,\\
        Department of Physics,\\
        Darmstadt University of Technology,\\
        Hochschulstr. 6,\\      
        D-64289 Darmstadt, Germany\\
${}^\text{b}$ 
Department of Physics,\\
Michigan Technological University,\\
Houghton, MI 49931, USA
}

\date{\today}    
\maketitle

\begin{abstract}
The purpose of this paper is the fundamental theory of the non-uniform
motion of dislocations in two and three space-dimensions.
We investigate the non-uniform motion of an arbitrary distribution of
dislocations, a dislocation loop and straight dislocations
in infinite media using the theory of incompatible elastodynamics.
The equations of motion are derived for non-uniformly moving dislocations. 
The retarded elastic fields produced by a distribution of dislocations
and the retarded dislocation tensor potentials are determined. 
New fundamental key-formulae for the dynamics of dislocations are derived
(Jefimenko type and Heaviside-Feynman type equations of dislocations).
In addition, exact closed-form solutions of the elastic fields 
produced by a dislocation loop are calculated as retarded line integral expressions for subsonic motion. 
The fields of the elastic velocity and elastic distortion surrounding the 
arbitrarily moving dislocation loop
are given explicitly in terms of the so-called 
three-dimensional elastodynamic Li\'enard-Wiechert tensor potentials. 
The two-dimensional elastodynamic Li\'enard-Wiechert tensor potentials
and the near-field approximation of the elastic fields
for straight dislocations
are calculated.
The singularities of the near-fields of accelerating screw and edge
dislocations are determined.
\\

\end{abstract}
\section{Introduction}

The investigation of the non-uniform motion of dislocations 
is an important interdisciplinary research field.
It has attracted the attention of scientists from several 
different fields such as
applied mathematics, solid state physics, 
material science, continuum mechanics, and 
seismology (see, e.g, \citep{Lardner,AR,Pujol}).
Typical problems of elastodynamics of dislocations 
are the determination of elastic fields produced by the non-uniform
motion of straight dislocations and dislocation loops.

The theory of the non-uniform motion of Volterra dislocations has a long history 
starting with the famous and well-known papers of~\citet{Eshelby51,Eshelby53}.
Using an electromagnetic analogy, \citet{Eshelby53}
found the elastic fields 
of a non-uniformly moving screw dislocation.
The problem of the non-uniform motion of a gliding 
edge dislocation was solved by~\citet{Mura64}.
\citet{Mura64} gave the solution in terms 
of `stress (or potential) functions' 
for the velocity  and elastic 
distortion fields (see also \citep{WW80,Lardner}). 
\citet{Lazar2011} has written a systematic 
review paper about the non-uniform motion of straight dislocations including
the solution of a non-uniformly climbing edge dislocation.

The behaviour of a 
straight dislocation is somehow particular, because at any time
the fields are determined not only by the instantaneous values, 
but also by the values in the past~\citep{Eshelby51,Eshelby53}. 
As \citet{Eshelby51} succinctly put it: `The dislocation is haunted by 
its past'.
For that reason, 
all the solutions of the elastic fields of non-uniformly moving 
straight dislocations are given in the form as time integrals and show an
afterglow. 
Due to the afterglow, 
Huygens' principle is not valid in two dimensions~(see, e.g., ~\citep{Wl,Achenbach,Strauss}). 
In general, a two-dimensional wave-motion possesses a `tail'.
The wave motion in two-dimensions and the afterglow effect are discussed more 
in detail by~\citet{Baker,Barton}, and \citet{Lazar2011}.

The non-uniform motion of dislocation loops was also studied.
The solution of non-uniformly moving dislocation loops was first 
formulated by~\citet{Mura63} in terms of the three-dimensional 
elastodynamic Green tensor 
as double integrals over the loop curve and time (see also~\citep{Mura}).  
The mathematical formulation of moving dislocation loops 
and analogous double integral presentations were 
also given by~\citet{Kossecka69} and \citet{Kossecka77,Kossecka77b}.
It was pointed out by~\citet{XM83} (see also~\citep{XM82b,XM84}) 
that some care is necessary in the calculation of the elastic fields 
produced by non-uniformly moving dislocations 
in order to avoid non-integrable singularities. 
\citet{XM83} showed that the general expressions for the 
velocity and elastic distortion fields of dislocations 
given by \citet{Mura63,Mura} 
are not free of non-integrable singularities. 
The reason is 
that the integration and differentiation cannot be changed in some cases.

In general,
the problem of moving dislocation loops is a three-dimensional problem in space.
It is well-known that 
Huygens' principle is only valid for odd space dimensions: $N\ge 3$
\citep{Wl,Achenbach,Strauss}.
For that reason,
there is no afterglow in three dimensions.
In standard books on electrodynamics (see, e.g., \citep{LL,Jackson}), 
the electric scalar and magnetic vector potentials 
of a non-uniformly moving point charge
which are the famous Li\'enard-Wiechert potentials~\citep{Lienard,Wiechert} 
can be found. 
It is quite surprising 
that nothing has been done in this direction in the elastodynamics of
moving dislocation loops up to now. 
In elastodynamics, 
only the retarded potentials were given
for the waves produced by body forces,
using the Helmholtz decomposition (see, e.g., \citep{Achenbach,Miklowitz}).
A more general expression for the retarded potential in elastodynamics was given 
by~\citet{Hudson}.
Though Emil Wiechert was a director of the geophysical laboratory in
G\"ottingen, he formulated his theory for electrodynamics~\citep{Wiechert}, but not
for elastodynamics, with which he was certainly familiar.
In this paper, we investigate 
some fundamental problems
of the elastodynamical theory of dislocations in analogy to the electromagnetic 
field theory.  
Especially, we want to develop the Li\'enard-Wiechert potentials
for the elastodynamics and to apply them to dislocation loops and straight 
dislocations.

This paper is organized as follows.
In Section~2, the framework of incompatible elastodynamics and 
the equations of motion of dislocations are presented.
In Section~3, the equations of motion for an 
arbitrary three-dimensional distribution of dislocations are solved, 
using the three-dimensional elastodynamic Green tensor.
As a result the retarded elastic fields are given.
The non-uniform motion of a closed dislocation loop is studied 
in Section~4. 
The so-called elastodynamic Li\'enard-Wiechert tensor potentials are determined.
The elastic fields are given in terms of so-called 
Li\'enard-Wiechert tensor potentials.
The static limit of the elastic fields of the non-uniformly moving dislocation
loop  is given in Section~5.
In Section~6, the two-dimensional 
elastodynamic Li\'enard-Wiechert tensor potentials and the elastic fields
of non-uniformly moving straight dislocations are calculated.
In Section~7, the near-field approximation of accelerating screw
and edge dislocations are calculated. 
In a straightforward manner, the $1/R$-singularity
and a logarithmic singularity associated with the acceleration of the 
dislocation are found.
The relation between the elastodynamic Li\'enard-Wiechert tensor potentials
and Mura's dislocation tensor potentials is presented in 
Section~8. The retarded dislocation tensor potentials
and the proper elastodynamic Li\'enard-Wiechert tensor potentials
of a dislocation loop are determined.
In Section~9, the conclusions are given.

\section{The equations of motion of dislocations}

In this section, we derive the equations of motion for dislocations in the framework
of incompatible elastodynamics~(see, e.g., 
\citep{Mura63,Mura,Kossecka69,Kosevich,LL2,Lazar2011}).
An unbounded, isotropic, homogeneous, linearly elastic solid is considered.
In the elasticity theory of self-stresses
the equilibrium condition is\footnote{We 
use the usual notation $\beta_{ij,k}:=\pd_k \beta_{ij}$ and 
$\dot{\beta}_{ij}:=\pd_t \beta_{ij}$.}
\begin{align}
\label{EC0}
\dot{p}_i =\sigma_{ij,j}\, ,
\end{align}
where $\Bp$ and $\Bsigma$ are the linear momentum vector and the force
stress tensor, respectively. 
In the incompatible linear elasticity,  
the momentum vector $\Bp$ and the stress tensor 
$\Bsigma$ can be expressed in terms of the incompatible 
elastic velocity (particle velocity) vector $\Bv$ 
and the incompatible elastic distortion tensor $\Bbeta$ by means of the two
constitutive relations
\begin{align}
\label{CR-p}
p_i&= \rho\,  v_i\,,\\
\label{CR-t}
\sigma_{ij}&=C_{ijkl}\, \beta_{kl}\, ,
\end{align}
where $\rho$ denotes the mass density and 
$C_{ijkl}$ is the tensor of elastic moduli.
The tensor $C_{ijkl}$ is characterized by the symmetry properties
\begin{align}
C_{ijkl}=C_{jikl}=C_{ijlk}=C_{klij}\, .
\end{align}
If the constitutive relations~(\ref{CR-p}) and (\ref{CR-t}) are substituted in 
Eq.~(\ref{EC0}), the equilibrium condition expressed in terms of 
the elastic fields $\Bv$ and $\Bbeta$ is obtained
\begin{align}
\label{EC}
\rho\, \dot{v}_i =C_{ijkl}\beta_{kl,j}\, .
\end{align}
The presence of dislocations makes the elastic fields incompatible 
which means that they are not anymore simple gradients or time derivatives 
of a displacement vector $\Bu$.
Unlike the displacement field and the plastic fields, the elastic
fields are physical state quantities of dislocations. 
This is one reason, why we deal only with the calculation of the elastic fields in this paper.

Other important tensor fields of dislocations are the dislocation density and 
dislocation current tensors (e.g.~\citep{Kosevich,Lazar2011b}).
The dislocation density tensor $\BT$ and the dislocation current tensor $\BI$ 
are defined by (see also~\citep{Bovet,Lazar2011b})
\begin{align}
\label{T}
T_{ijk}&=\beta_{ik,j}-\beta_{ij,k}\,,\\
\label{I}
I_{ij}&=\dot{\beta}_{ij}-v_{i,j}\,. 
\end{align}
The dislocation current tensor $\BI$ was originally introduced 
by~\citet{Kosevich62,Kosevich65} and \citet{Hollaender60,Hollaender62} 
(see also~\citep{Kosevich,LL2,Teodosiu70,goleb}).
The dislocation current tensor~(\ref{I}) is the difference of 
two pieces: the time derivative of the elastic distortion, and 
the elastic velocity gradient.
It is noted that $T_{ijk}=-T_{ikj}$.
Both $\BT$ and $\BI$ have nine independent components.
A field theoretical justification for the structure of $\BT$ and $\BI$ was
given by~\citet{LA08} (see also~\citep{Lazar2011b}).
Moreover, they fulfill the Bianchi identities 
(see also~\citep{Guenther67,Guenther73})
\begin{align}
\label{BI1}
\epsilon_{jkl}\,T_{ijk,l}&=0\, ,\\
\label{BI2}
\dot{T}_{ijk} + I_{ij,k}-I_{ik,j}&= 0\, ,    
\end{align}
which are `conservation' laws. 
Here $\epsilon_{jkl}$ denotes the Levi-Civita tensor.
Eq.~(\ref{BI1}) states that dislocations cannot end inside the medium and
Eq.~(\ref{BI2}) means that the time evolution of the dislocation density tensor
$\BT$ is determined by the `curl' of the dislocation current tensor $\BI$.
The tensors $\BT$ and $\BI$ may describe single straight dislocations, 
dislocation loops and an arbitrary distribution of dislocations.

Alternatively, we may rewrite the tensor $\BT$ as (dual) tensor of rank two
\begin{align}
\label{alpha}
\alpha_{ij}=\frac{1}{2}\,\epsilon_{jkl} T_{ikl}=\epsilon_{jkl}\beta_{il,k}
\end{align}
with the inverse relation
\begin{align}
\label{alpha-inv}
T_{ikl}=\epsilon_{jkl}\alpha_{ij}\,.
\end{align}
The tensor $\Balpha$ is the usual dislocation density 
tensor\footnote{
In the literature, the notations of the dislocation density tensor and 
the dislocation current tensor are not unique:
$\Balpha(\text{\citet{Lazar2011b}})
=\Balpha(\text{\citet{Kossecka69,Kossecka74}})
=\Balpha^\TT(\text{\citet{Kossecka77}})
=\Balpha^\TT(\text{\citet{Kroener58}})
=\Balpha^\TT(\text{\citet{deWit2}})
=-\Balpha(\text{\citet{Teodosiu70}})
=-\Balpha^\TT(\text{\citet{Kosevich})}
$
and
$\BI(\text{\citet{Lazar2011b}})
=\BI(\text{\citet{Kossecka69}})
=-\BI^\TT(\text{\citet{Kossecka77}})
=-\BI(\text{\citet{Teodosiu70}})
=\BI^\TT(\text{\citet{Kosevich})}
$. }.
Then the Bianchi identities (conservation laws)~(\ref{BI1}) and (\ref{BI2})
simplify to (see also~\citep{Kosevich,Kossecka75,LL2,Schaefer,Teodosiu70})
\begin{align}
\label{BI1-b}
\alpha_{ij,j}&=0\, ,\\
\label{BI2-b}
\dot{\alpha}_{ij} + \epsilon_{jkl}I_{ik,l}&= 0\, .
\end{align}

Now we derive separated equations for the elastic fields $\Bbeta$ and $\Bv$
as equations of motion.
If we differentiate the Eq.~(\ref{EC}) with respect to $x_{m}$ and use
Eqs.~(\ref{T}) and (\ref{I}) to eliminate $\Bv$,
we get the equation of motion for 
the incompatible elastic distortion tensor $\boldsymbol{\beta}$ 
(see also~\citep{Kossecka69,Teodosiu70,Lazar2011})
\begin{align}
\label{B-NE}
\rho\, \ddot{\beta}_{im}- C_{ijkl}\,\beta_{km,jl}=C_{ijkl}\,T_{kml,j}+\rho\dot{I}_{im}\, ,
\end{align}
where the dislocation
density and the dislocation current tensors are the sources.
Eq.~(\ref{B-NE}) is a tensorial Navier equation for $\boldsymbol{\beta}$.
Similarly, if we perform the differentiation of Eq.~(\ref{EC}) 
with respect to time and use Eq.~(\ref{I}) to eliminate $\Bbeta$, we obtain
the   equation of motion for
the incompatible elastic velocity vector $\boldsymbol{v}$
(see also~\citep{Kossecka69,Teodosiu70,Kosevich,Lazar2011})
\begin{align}
\label{v-NE}
\rho\,\ddot{v}_i-C_{ijkl}\, v_{k,jl}=C_{ijkl}\,I_{kl,j}\, ,
\end{align}
where the dislocation current tensor is the source term.
Eq.~(\ref{v-NE}) is a vectorial Navier equation for $\boldsymbol{v}$.
We want to note that also \citet{Rogula} obtained equations of the form
(\ref{B-NE}) and (\ref{v-NE}) for dislocations in a pseudo-continuum

\section{The retarded elastic fields}
In this section, we calculate the retarded elastic fields produced by a
three-dimensional distribution of dislocations.
The solutions of Eqs.~(\ref{B-NE}) and (\ref{v-NE}) can be represented as
convolution integrals~\citep{Mura63,Kossecka69,Kossecka77b,Lazar2011}.
For an unbounded medium and under the assumption of zero 
initial conditions,
which means that $\Bbeta(\rr,t_0)$ and $\Bv(\rr,t_0)$ 
and their first time derivatives are zero 
for $t_0\rightarrow-\infty$, the solutions of $\Bbeta$ and $\Bv$ 
can be represented as
\begin{align}
\beta_{im}(\rr,t)&=
\label{B-M}
\pd_k
\int_{-\infty}^t \int_{-\infty}^\infty
C_{jkln}\, G_{ij}(\rr-\rr', t-t')\, T_{lmn}(\rr',t')\, \d \rr'\, \d t'
\nonumber\\
&\quad
+\pd_t \int_{-\infty}^t \int_{-\infty}^\infty
\rho\, G_{ij}(\rr-\rr', t-t')\, {I}_{jm}(\rr',t')\, \d \rr'\, \d t'
\end{align}
and
\begin{align}
v_i(\rr,t)=
\label{v-M}
\pd_k\int_{-\infty}^t \int_{-\infty}^\infty
C_{jklm}\, G_{ij}(\rr-\rr', t-t')\, I_{lm}(\rr',t')\, \d \rr'\, \d t'\, .
\end{align}
It is obvious that dislocations act as a source of the elastodynamic fields.
A proof that Eqs.~(\ref{B-M}) and (\ref{v-M}) fulfill Eq.~(\ref{EC})
can be found in~\citet{Kossecka75}.
Here, $G_{ij}$ is the elastodynamic Green tensor of the anisotropic Navier equation 
defined by
\begin{align}
\label{GF-e}
\big[\delta_{ik}\,\rho \, \pd_{tt}-C_{ijkl}\pd_j\pd_l\big] G_{km}=\delta_{im}\,
\delta(t)\delta(\rr)\, ,
\end{align}
where $\delta(.)$ denotes the Dirac delta function and
$\delta_{ij}$ is the Kronecker delta.
For isotropic materials, the tensor of elastic moduli reduces to 
\begin{align}
\label{C}
C_{ijkl}=\lambda\, \delta_{ij}\delta_{kl}
+\mu\big(\delta_{ik}\delta_{jl}+\delta_{il}\delta_{jk})\, ,
\end{align}
where $\lambda$ and $\mu$ are the Lam{\'e} constants.
Substituting Eq.~(\ref{C}) in Eq.~(\ref{GF-e}), 
the isotropic Navier equation for the dynamic Green tensor was obtained
\begin{align}
\big[\delta_{ik}\,\rho \, \pd_{tt}- \delta_{ik}\, \mu\, \Delta
-(\lambda+\mu)\, \pd_i \pd_k \big] G_{km}=\delta_{im}\,
\delta(t)\delta(\rr)\, ,
\end{align}
where $\Delta$ denotes the Laplacian.
When the material is isotropic and infinitely extended, 
the three-dimensional elastodynamic Green tensor reads~\citep{Love,Eringen75,AR,Pujol}
\begin{align}
\label{GT}
G_{ij}(\rr,t)&=\frac{1}{4\pi\rho }\, 
\Bigg\{
\frac{\delta_{ij}}{r c^2_\TT}\, \delta(t-r/c_\TT)
+\frac{x_i x_j}{r^3}\,
\bigg(
\frac{1}{c^2_\TL}\,  \delta(t-r/c_\TL)
-\frac{1}{c^2_\TT}\,  \delta(t-r/c_\TT)\bigg)\nonumber\\
&\hspace{15mm}
+\bigg(\frac{3x_i x_j}{r^2}-\delta_{ij}\bigg)
\frac{1}{r^3}\,\int_{r/c_\TL}^{r/c_\TT}\tau\,\delta(t-\tau)\,\d \tau
\Bigg\}\,,
\end{align}
where $r=\sqrt{x^2+y^2+z^2}$.
It should be pointed out that Eq.~(\ref{GT}) is the retarded Green tensor.
Here, $c_\TL$ and $c_\TT$ denote the velocities of the 
longitudinal and transversal elastic waves (sometimes called P- and S-waves). 
The two sound velocities can be given in terms of the Lam\'e constants ($c_\TT<c_\TL$)
\begin{align}
\label{c}
c_{\TL}=\sqrt{\frac{2\mu+\lambda}{\rho}}\,,\qquad
c_{\TT}=\sqrt{\frac{\mu}{\rho}}\, .
\end{align}
The elastodynamic Green tensor~(\ref{GT}) is a tensor-valued distribution with support
along the two sound cones $r=c_\TT t$ and $r=c_\TL t$ 
as well as in the region between them. 
Eq.~(\ref{GT}) consists of near-field and far-field terms.
The first terms in (\ref{GT}) decay as $1/r$ and thus, they are the far-field terms. 
The last term in (\ref{GT}) decays more rapidly like $1/r^2$ which gives 
the near-field term (see, e.g., \citep{Pujol}).

If we use the relation
\begin{align}
\label{Rel}
\frac{1}{r^2}\,\int_{r/c_\TL}^{r/c_\TT}\tau\,\delta(t-\tau)\,\d \tau=
\int_{1/c_\TL}^{1/c_\TT}\kappa\,\delta(t-\kappa r)\,\d \kappa\, ,
\end{align}
substitute the Green tensor~(\ref{GT}) into Eqs.~(\ref{B-M}) and
(\ref{v-M}) 
and perform the integration in $t'$, 
the retarded elastic fields produced by an arbitrary
three-dimensional distribution of dislocations are found
\begin{align}
\beta_{im}(\rr,t)&=
\label{B-M-ret}
\frac{1}{4\pi\rho}\, \pd_k\int_\VV
C_{jkln}\, 
\bigg\{
\frac{1}{c^2_\TT}\bigg(\frac{\delta_{ij}}{R}
-\frac{R_i R_j}{R^3}\bigg) T_{lmn}(\rr',t_\TT)
+\frac{1}{c^2_\TL}\,\frac{R_i R_j}{R^3}\,T_{lmn}(\rr',t_\TL)
\nonumber\\
&\hspace{16mm}
+\bigg(\frac{3R_i R_j}{R^3}-\frac{\delta_{ij}}{R}\bigg)
\int_{1/c_\TL}^{1/c_\TT} \kappa\,T_{lmn}(\rr',t_\kappa)\, \d \kappa\bigg\}\,
\d \rr'\nonumber\\
&\quad
+\frac{1}{4\pi}\, \pd_t\int_\VV
\bigg\{
\frac{1}{c^2_\TT}\bigg(\frac{\delta_{ij}}{R}
-\frac{R_i R_j}{R^3}\bigg) I_{jm}(\rr',t_\TT)
+\frac{1}{c^2_\TL}\,\frac{R_i R_j}{R^3}\,I_{jm}(\rr',t_\TL)
\nonumber\\
&\hspace{16mm}
+\bigg(\frac{3R_i R_j}{R^3}-\frac{\delta_{ij}}{R}\bigg)
\int_{1/c_\TL}^{1/c_\TT} \kappa\,I_{jm}(\rr',t_\kappa)\, \d \kappa\bigg\}\,
\d \rr'
\end{align}
and
\begin{align}
v_i(\rr,t)&=
\label{v-M-ret}
\frac{1}{4\pi\rho}\, \pd_k\int_\VV
C_{jklm}\, 
\bigg\{
\frac{1}{c^2_\TT}\bigg(\frac{\delta_{ij}}{R}
-\frac{R_i R_j}{R^3}\bigg) I_{lm}(\rr',t_\TT)
+\frac{1}{c^2_\TL}\,\frac{R_i R_j}{R^3}\,I_{lm}(\rr',t_\TL)
\nonumber\\
&\hspace{16mm}
+\bigg(\frac{3R_i R_j}{R^3}-\frac{\delta_{ij}}{R}\bigg)
\int_{1/c_\TL}^{1/c_\TT} \kappa\, I_{lm}(\rr',t_\kappa)\, \d \kappa
\bigg\}\,\d \rr'\, ,
\end{align}
where the so-called retarded times are given by
\begin{align}
\label{tT}
t_\TT&=t-\frac{R}{c_\TT}\, ,\\
\label{tL}
t_\TL&=t-\frac{R}{c_\TL}\, ,\\
\label{tkappa}
t_\kappa&=t-\kappa R\, ,
\end{align}
and $\kappa$ is a dummy variable with the dimension
$1/[\text{velocity}]$.
Here 
$t_\TT$ and $t_\TL$ are the transversal retarded time and
the longitudinal retarded time, respectively.
The retarded time $t_\kappa$ is an effective retarded time for the 
$\kappa$-integration with the limits $(1/c_\TL,1/c_\TT)$.
Since $c_\TL>c_\TT$, the retarded times fulfill:
$t_\TT>t_\TL$ and $t_\kappa\in[t_\TL,t_\TT]$.
Because the integrals~(\ref{B-M-ret}) and (\ref{v-M-ret}) are evaluated at the retarded
times, they are called retarded elastic fields.
Here $R=|\rr-\rr'|$ is the distance from the source point $\rr'$ to
the field point $\rr$ and is independent of $t$. 
The retarded elastic fields~(\ref{B-M-ret}) and (\ref{v-M-ret}) are just
integrals in $\rr'$. 
Here $\VV$ is a volume integral in $\R^3$.
The sources $\BT$ and $\BI$ at the position $\rr'$ depend on the retarded times.
The retarded elastic fields at the position $\rr$ and time $t$ contain
contributions from the past sound cones.
Elastodynamic fields and waves propagate with finite velocities.
Thus, there always is a time delay before a change in elastodynamic conditions 
initiated at a point of space can produce an effect at any other point of
space. This time delay is called elastodynamic retardation.
The retarded elastic fields~(\ref{B-M-ret}) and (\ref{v-M-ret}) consist of 
three characteristic parts. The first term is the transversal one, 
transmitting with speed $c_\TT$, and it corresponds to $S$-wave motion.
The second term is the longitudinal one, 
transmitting with speed $c_\TL$, and it corresponds to $P$-wave motion.
Finally, the third term is 
neither longitudinal nor transversal and it gives 
contribution arriving at the speeds between the two characteristic ones, 
which shows that this factor 
represents a combination of $P$-wave and $S$-wave motion.
For a three-dimensional distribution of dislocations, we conclude that
dislocations are retarded but not haunted by its past in
contrast to a straight dislocation.

Now we carry out the differentiations in Eqs.~(\ref{B-M-ret}) and 
(\ref{v-M-ret})
and use the relations
\begin{align}
&\pd_k T_{lmn}(\rr',t_{\text{ret}})=-\frac{R_k}{c R}\, \pd_t
T_{lmn}(\rr',t_{\text{ret}})\,,\quad
\pd_k I_{lm}(\rr',t_{\text{ret}})=-\frac{R_k}{c R}\, \pd_t
I_{lm}(\rr',t_{\text{ret}})\,,\nonumber\\
&{\text{with}}\quad 
t_{\text{ret}}=t_\TT,t_\TL,t_\kappa,\qquad
c=c_\TT,c_\TL,1/\kappa\,,
\end{align}
in order to find the retarded elastic field more explicitly.
We obtain
\begin{align}
&\beta_{im}(\rr,t)=
\label{B-ret-Jef}
-\frac{1}{4\pi\rho}\, C_{jkln}\, 
\int_\VV
\bigg\{
\frac{1}{c^2_\TT}\bigg(
\frac{\delta_{ij} R_k +\delta_{jk} R_i +\delta_{ik} R_j}{R^3}
-\frac{3R_iR_jR_k}{R^5}\bigg)T_{lmn}(\rr',t_\TT)\nonumber\\
&\qquad
+\frac{1}{c^3_\TT}\bigg(\delta_{ij}-\frac{R_i R_j}{R^2}\bigg) 
\frac{R_k}{R^2}\,\pd_t T_{lmn}(\rr',t_\TT)
-\frac{1}{c^2_\TL}\bigg(
\frac{\delta_{jk} R_i +\delta_{ik} R_j}{R^3}
-\frac{3R_iR_jR_k}{R^5}\bigg)T_{lmn}(\rr',t_\TL)\nonumber\\
&\quad
+\frac{1}{c^3_\TL}\,\frac{R_i R_jR_k }{R^4}\, \pd_t T_{lmn}(\rr',t_\TL)
-\bigg(\frac{\delta_{ij} R_k +3\delta_{jk} R_i +3\delta_{ik} R_j}{R^3}
-\frac{9R_iR_jR_k}{R^5}\bigg)
\int_{1/c_\TL}^{1/c_\TT} \kappa\, T_{lmn}(\rr',t_\kappa)\, \d \kappa
\nonumber\\
&\hspace{16mm}
+\bigg(\frac{3R_i R_j}{R^2}-\delta_{ij}\bigg)\frac{R_k}{R^2}
\int_{1/c_\TL}^{1/c_\TT} \kappa^2\, \pd_t T_{lmn}(\rr',t_\kappa)\, \d \kappa
\bigg\}\,\d \rr'\, \nonumber\\
&\hspace{16mm}
+\frac{1}{4\pi}\, \int_\VV
\bigg\{
\frac{1}{c^2_\TT}\bigg(\frac{\delta_{ij}}{R}
-\frac{R_i R_j}{R^3}\bigg)\, \pd_t I_{jm}(\rr',t_\TT)
+\frac{1}{c^2_\TL}\,\frac{R_i R_j}{R^3}\, \pd_t I_{jm}(\rr',t_\TL)
\nonumber\\
&\hspace{16mm}
+\bigg(\frac{3R_i R_j}{R^3}-\frac{\delta_{ij}}{R}\bigg)
\int_{1/c_\TL}^{1/c_\TT} \kappa\, \pd_t 
I_{jm}(\rr',t_\kappa)\, \d \kappa\bigg\}\,
\d \rr'
\end{align}
and
\begin{align}
&v_i(\rr,t)=
\label{v-ret-Jef}
-\frac{1}{4\pi\rho}\, 
C_{jklm}
\int_\VV 
\bigg\{
\frac{1}{c^2_\TT}\bigg(
\frac{\delta_{ij} R_k +\delta_{jk} R_i +\delta_{ik} R_j}{R^3}
-\frac{3R_iR_jR_k}{R^5}\bigg)I_{lm}(\rr',t_\TT)\nonumber\\
&\qquad
+\frac{1}{c^3_\TT}\bigg(\delta_{ij}-\frac{R_i R_j}{R^2}\bigg) 
\frac{R_k}{R^2}\,\pd_t I_{lm}(\rr',t_\TT)
-\frac{1}{c^2_\TL}\bigg(
\frac{\delta_{jk} R_i +\delta_{ik} R_j}{R^3}
-\frac{3R_iR_jR_k}{R^5}\bigg)I_{lm}(\rr',t_\TL)\nonumber\\
&\quad
+\frac{1}{c^3_\TL}\,\frac{R_i R_jR_k }{R^4}\, \pd_t I_{lm}(\rr',t_\TL)
-\bigg(\frac{\delta_{ij} R_k +3\delta_{jk} R_i +3\delta_{ik} R_j}{R^3}
-\frac{9R_iR_jR_k}{R^5}\bigg)
\int_{1/c_\TL}^{1/c_\TT} \kappa\, I_{lm}(\rr',t_\kappa)\, \d \kappa
\nonumber\\
&\hspace{16mm}
+\bigg(\frac{3R_i R_j}{R^2}-\delta_{ij}\bigg)\frac{R_k}{R^2}
\int_{1/c_\TL}^{1/c_\TT} \kappa^2\, \pd_t I_{lm}(\rr',t_\kappa)\, \d \kappa
\bigg\}\,\d \rr'\, .
\end{align}
These are the general 
retarded elastic fields produced by a time-dependent dislocation distribution.
We now see from Eqs.~(\ref{B-ret-Jef}) and (\ref{v-ret-Jef})
that the elastic distortion has three sources: the dislocation density $\BT$,
the time derivative of $\BT$, and the time derivative of $\BI$. 
And  we see that the elastic velocity has two sources: the dislocation current
$\BI$, and the time derivative of $\BI$. All of these sources are retarded due to
the retarded times.
Equations~(\ref{B-ret-Jef}) and (\ref{v-ret-Jef})
show that, since both equations 
contain the time derivative of $\BI$, the elastic distortion and the elastic 
velocity fields are created by the same
time-variable dislocation current depending on the retarded times. 
Thus, the time-changing of the dislocation current $\BI$ is the common source
for $\Bbeta$ and $\Bv$.
The three $\delta_{ij}$-terms multiplied by $1/c_\TT$-factors 
in Eq.~(\ref{B-ret-Jef}) and the two $\delta_{ij}$-terms multiplied 
by $1/c_\TT$-factors in Eq.~(\ref{v-ret-Jef})
are analogous to the Jefimenko formulae for the electromagnetic
field strengths (see~\citep{Jefimenko,Griffiths,Jackson,HM}).
Originally, \citet{Jefimenko} (see also \citep{Griffiths}) derived
the proper time-dependent generalizations of the Coulomb law and 
the Biot-Savart law as causal solutions of the Maxwell equations\footnote{
In electrodynamics, 
the Jefimenko formulae for the electric field strength $\BE$ and 
the magnetic field strength $\BB$ are given by~\citep{Griffiths,Jackson}:
\begin{align*}
\BE(\rr,t)&=
\frac{1}{4\pi \epsilon_0}
\int_\VV \bigg(
\frac{\rho(\rr',t-R/c)}{R^3}\,\RR
+\frac{\pd_t \rho(\rr',t-R/c)}{c R^2}\,\RR
-\frac{\pd_t \BJ(\rr',t-R/c)}{c^2 R}\bigg)
 \d \rr'\,,\\
\BB(\rr,t)&=
\frac{1}{4\pi \epsilon_0 c^2}
\int_\VV \bigg(
\frac{\BJ(\rr',t-R/c)}{R^3}
+\frac{\pd_t \BJ(\rr',t-R/c)}{c R^2}\bigg)\times \RR\ 
 \d \rr'\,,
\end{align*}
where $\rho$ is the electric charge density, $\BJ$ 
denotes the electric current density vector, $c$ denotes the speed of light
and $\epsilon_0$ is the permittivity of vacuum.}.
It is obvious that the retarded elastic fields~(\ref{B-ret-Jef}) and
(\ref{v-ret-Jef}) are more complicated than the Jefimenko equations in 
electrodynamics due to the tensor structure of the elastodynamical
Green tensor and the dislocation-source fields.
Nevertheless, we may call Eqs.~(\ref{B-ret-Jef}) and (\ref{v-ret-Jef})
the Jefimenko type equations of dislocations.
The causal dependencies of the elastodynamic phenomena 
of the motion of dislocations
are described by
Eqs.~(\ref{B-ret-Jef}) and ~(\ref{v-ret-Jef}) which are exact solutions
of the Navier equations~(\ref{B-NE}) and (\ref{v-NE}) involving integrals over
retarded dislocation sources.

Performing the $\kappa$-integration and using Eqs.~(\ref{C}) and (\ref{c})
the static limit of Eq.~(\ref{B-ret-Jef}),
which gives the (static) dislocation version of the Biot-Savart law is calculated as
\begin{align}
\label{B-Jef-stat}
\beta_{im}(\rr)
=-\frac{1}{8\pi(1-\nu)}\int_\VV
\bigg((1-2\nu)\frac{\delta_{il}R_n+\delta_{in}R_l-\delta_{ln}R_i}{R^3}
+\frac{3R_iR_lR_n}{R^5}\bigg) T_{lmn}(\rr')\, \d \rr'\,.
\end{align}
It gives the correct expression for the 
elastic distortion tensor in terms of the dislocation density tensor~$\BT$.
Eq.~(\ref{B-Jef-stat}) 
is the explicite expression 
of the isotropic version of the so-called Mura-Willis formula
(see, e.g.,~\citep{deWit2,deWit3}).
Here $\nu$ is the Poisson ratio with
$\lambda=2\mu\nu/(1-2\nu)$ and $\nu=\lambda/[2(\lambda+\mu)]$.

\section{A non-uniformly moving dislocation loop}
Investigating the non-uniform motion of a dislocation loop, 
we consider a closed loop of arbitrary shape (planar or non-planar) 
that moves arbitrary.
The dislocation density tensor and the dislocation current tensor of 
a dislocation loop at the position $\Bs(t)$ are represented by line integrals of
the form (e.g. \citep{Kossecka69,Kossecka75,Kossecka77})
\begin{align}
\label{T-L}
T_{ijk}(\rr,t)&=b_i \epsilon_{jkl}\oint_{L(t)}\delta(\rr-\Bs(t))\, \d
L_l(\Bs(t))\,,\\
\label{I-L}
I_{ij}(\rr,t)&=b_i \epsilon_{jkl}\oint_{L(t)}V_k(t)\, \delta(\rr-\Bs(t))\, \d
L_l(\Bs(t))\,,
\end{align}
where $\BV=\dot{\Bs}$ denotes the velocity of the dislocation loop
at any point $\Bs(t)$ on the loop, $b_i$ is the Burgers vector, 
$L(t)$ is the dislocation loop curve at time $t$ and $\d L_l$
is a line element along the loop.
Since $L(t)$ may change its position with time $t$,
it has a more complicated structure than in the static case.
$L(t)$ is the collection of all points on the dislocation line.
Only subsonic source-speeds will be admitted ($|\BV|<c_\TT$).
If we substitute Eqs.~(\ref{T-L}) and (\ref{I-L}) into (\ref{B-M}) and 
(\ref{v-M}) and after the integration in $\rr'$, we obtain for 
the elastic fields of a dislocation loop
\begin{align}
\beta_{im}(\rr,t)&=
\label{B-M-L}
\pd_k
\int_{-\infty}^t \oint_{L(t')}
\epsilon_{mnp}\, C_{jkln}\, G_{ij}(\rr-\Bs(t'), t-t')\, 
b_l\,  \d L_p(\Bs(t'))\, \d t'
\nonumber\\
&\quad
+\pd_t \int_{-\infty}^t  \oint_{L(t')}
\rho\, G_{ij}(\rr-\Bs(t'), t-t')\, b_j \epsilon_{mnp} V_n \,  \d L_p(\Bs(t'))\, \d t'
\end{align}
and
\begin{align}
v_i(\rr,t)=
\label{v-M-L}
\pd_k\int_{-\infty}^t \oint_{L(t')}
C_{jklm}\, G_{ij}(\rr-\Bs(t'), t-t')\, 
 b_l \epsilon_{mnp} V_n \,  \d L_p(\Bs(t'))\, \d t'\, ,
\end{align}
which are in agreement with the formulae given by~\citet{XM83}.

Integration in the expressions~(\ref{B-M-L}) and
(\ref{v-M-L}) may be performed in time. 
In this way, 
we find the elastic fields as line integrals around the loop $L(t')$
\begin{align}
\beta_{im}(\rr,t)&=
\label{B-M-L2}
\pd_k \oint_{L(t')}
\epsilon_{mnp}\, C_{jkln}\, \phi_{ij}\, 
b_l\,  \d L_p(\Bs(t'))
+\pd_t \oint_{L(t')}
\rho\, A_{ijk}\, b_j \epsilon_{mkp} \,  \d L_p(\Bs(t'))\,
\end{align}
and
\begin{align}
v_i(\rr,t)=
\label{v-M-L2}
\pd_k \oint_{L(t')}
C_{jklm}\, A_{ijn}\, 
 b_l \epsilon_{mnp}\,  \d L_p(\Bs(t'))\,,
\end{align}
where we have introduced the elastodynamic 
Li\'enard-Wiechert tensor potentials $\phi_{ij}$ and $A_{ijk}$ 
in terms of the elastodynamic Green tensor 
$G_{ij}(\rr-\Bs(t'), t-t')$:
\begin{align}
\phi_{ij}(\rr,t)&=
\label{phi-0}
\int_{-\infty}^t\int_{-\infty}^\infty G_{ij}(\rr-\rr', t-t')\, 
\delta(\rr'-\Bs(t'))\, \d \rr'\, \d t'\nonumber\\
&=\int_{-\infty}^t G_{ij}(\rr-\Bs(t'), t-t')\, \d t'\, ,\\
A_{ijk}(\rr,t)&=
\label{A-0}
\int_{-\infty}^t\int_{-\infty}^\infty G_{ij}(\rr-\rr', t-t')\, V_k(t')\,
\delta(\rr'-\Bs(t'))\, \d \rr'\, \d t'\nonumber\\
&=\int_{-\infty}^t G_{ij}(\rr-\Bs(t'), t-t')\, V_k(t')\, \d t'\, .
\end{align}
More precisely, Eqs.~(\ref{phi-0}) and (\ref{A-0}) are the elastodynamical
Li\'enard-Wiechert tensor potentials corresponding to delta-point sources
acting on the position $\Bs(t')$ which have to be integrated over the loop line element $\d L_p(\Bs(t'))$ in 
Eqs.~(\ref{B-M-L2}) and (\ref{v-M-L2}).
The property of the Green tensor~(\ref{GF-e}) 
leads to the following wave equations
for the  elastodynamic Li\'enard-Wiechert tensor potentials~(\ref{phi-0}) 
and (\ref{A-0})
\begin{align}
\label{phi-pde}
&\big[\delta_{ik}\,\rho \, \pd_{tt}-C_{ijkl}\pd_j\pd_l\big] \phi_{km}=
\delta_{im}\,\delta(\rr-\Bs(t))\, ,\\
\label{A-pde}
&\big[\delta_{ik}\,\rho \, \pd_{tt}-C_{ijkl}\pd_j\pd_l\big] A_{kmn}=
\delta_{im}\, V_n (t)\,\delta(\rr-\Bs(t))\, ,
\end{align}
with Eq.~(\ref{C}).

Substituting the elastodynamic Green tensor (\ref{GT}) into Eqs.~(\ref{phi-0}) and
(\ref{A-0}), the integration in time may be performed. 
Before the integration $R=|\rr-\rr'|$ is a function of $\rr$ and $\rr'$;
after the integration, which fixes $\rr'=\Bs(t)$, $R=|\rr-\Bs(t)|$ is a function of
$\rr$ and $\Bs(t)$. 
The structure of the Green tensor~(\ref{GT}) 
produces three characteristic integrals 
which we have to calculate.
We express the integrals in terms of retarded variables by appeal to the 
relation~\citep{Roos,Jones,Barton}
\begin{align}
\label{Roos}
\int \delta(f(t'))\,g(t')\, \d t'=\frac{g(t')}{|\d f/\d t'|}\bigg|_{{\text{at}}\, f(t')=0}
\, .
\end{align}
Mathematically, the factor $1/|\d f/\d t'|$ is the Jacobian of the 
transformation from $t'$ to the new integration variable $f(t')$.
This mapping between the two variables is one-to-one if the Jacobian is different from zero. A sufficient condition for this is that the velocity of the source 
(dislocation) is less than the wave speed (subsonic motion).
The first integral can be carried out with
\begin{align}
\label{Int1}
\int \frac{\delta(t-t'-|\rr-\Bs(t')|/c_\TT)}{|\rr-\Bs(t')|}\, \d t'
=\frac{1}{R-\BR\cdot\BV/c_\TT}\bigg|_{t'=t_\TT}\,,
\end{align}
where $\BR=\rr-\Bs(t')$ and $\BV=\BV(t')$.
$\BR$ is the distance vector 
from the position of the source $\Bs$, the sender of
elastic waves, to the point of the observer $\rr$, the receiver of the
elastic waves.
The second integral is
\begin{align}
\label{Int2}
\int \frac{(\rr-\Bs(t'))_i (\rr-\Bs(t'))_j}{|\rr-\Bs(t')|^3}\, 
\delta(t-t'-|\rr-\Bs(t')|/c_{\TL,\TT})\, \d t'
=\frac{R_iR_j}{R^2}\,\frac{1}{R-\BR\cdot\BV/c_{\TL,\TT}}\bigg|_{t'=t_{\TL,\TT}}\, .
\end{align}
Using the relation~(\ref{Rel}),
we perform the third integral as follows
\begin{align}
\label{Int3}
&\int \bigg(\frac{3(\rr-\Bs(t'))_i (\rr-\Bs(t'))_j}{|\rr-\Bs(t')|^3}
-\frac{\delta_{ij}}{|\rr-\Bs(t')|}\bigg)
\int_{1/c_\TL}^{1/c_\TT}\kappa\,\delta(t-t'-\kappa |\rr-\Bs(t')|)\,\d \kappa\, \d t'
\nonumber\\
&\quad
=\int_{1/c_\TL}^{1/c_\TT}
\bigg(\frac{3R_i R_j}{R^2}-\delta_{ij}\bigg)
\frac{\kappa\,\d \kappa}{R-\kappa\, \BR\cdot\BV}\bigg|_{t'=t_\kappa}\, .
\end{align}
From the argument that the delta functions vanish
$f(t')=0$ in Eqs.~(\ref{Int1})--(\ref{Int3}), we obtain the condition
\begin{align}
\label{t-ret}
t-t'-|\rr-\Bs(t')|/c=0\,,\qquad{\text{with}}\qquad c=c_\TT,c_\TL,1/\kappa\,.
\end{align}
Unfortunately, the retarded times $t_c=t'(\rr,t)$ 
are not given directly,
and they must be determined by solving Eq.~(\ref{t-ret}) what can be quite
tedious. Only in some simple cases $t_c$ is easy to find.
If the dislocation loop is moving with subsonic speed, the solution of 
Eq.~(\ref{t-ret}) is unique.
The retarded times are a result of the finite speeds of 
propagation for elastodynamic waves.
In Eqs.~(\ref{Int1})--(\ref{Int3}) we have used the relation
\begin{align}
\bigg|\frac{\d f(t')}{\d t'}\bigg|_{t'=t_{\text{ret}}}=
1-\frac{\BR\cdot\BV}{c\, R}\bigg|_{t'=t_{\text{ret}}}>0\qquad
\text{for}\ |\BV|<c_\TT\,,\quad c=c_\TT, c_\TL, 1/\kappa\quad
{\text{and}}\  t_{\text{ret}}=t_\TT, t_\TL, t_\kappa\, ,
\end{align}
where $f(t')=t-t'- |\rr-\Bs(t')|/c$.

Carrying out the $t'$-integration in Eqs.~(\ref{phi-0}) and (\ref{A-0}),
we find the explicit expressions for the elastodynamic 
Li\'enard-Wiechert tensor potentials of a `point dislocation source' 
acting on the position $\Bs(t')$
\begin{align}
\label{phi}
4\pi\rho\, \phi_{ij}(\rr,t)&=
\frac{1}{c^2_\TT}
\bigg[\bigg(\delta_{ij}-\frac{R_iR_j}{R^2}\bigg)
 \frac{1}{R-\BR\cdot\BV/c_\TT}\bigg]\bigg|_{t'=t_\TT}
+\frac{1}{c^2_\TL}\,
\bigg[\frac{R_iR_j}{R^2}\,
 \frac{1}{R-\BR\cdot\BV/c_\TL}\bigg]\bigg|_{t'=t_\TL}
\nonumber\\
&\qquad
+\int_{1/c_\TL}^{1/c_\TT}
\bigg[\bigg(\frac{3R_i R_j}{R^2}-\delta_{ij}\bigg)
\frac{\kappa\,\d \kappa}{R-\kappa\, \BR\cdot\BV}\bigg]\bigg|_{t'=t_\kappa}
\\
\label{A}
4\pi\rho\, A_{ijk}(\rr,t)&=
\frac{1}{c^2_\TT}
\bigg[\bigg(\delta_{ij}-\frac{R_iR_j}{R^2}\bigg)
\frac{V_k}{R-\BR\cdot\BV/c_\TT}\bigg]\bigg|_{t'=t_\TT}
+\frac{1}{c^2_\TL}\, 
\bigg[\frac{R_iR_j}{R^2}\,\frac{V_k}{R-\BR\cdot\BV/c_\TL}\bigg]\bigg|_{t'=t_\TL}
\nonumber\\
&\qquad
+\int_{1/c_\TL}^{1/c_\TT}
\bigg[\bigg(\frac{3R_i R_j}{R^2}-\delta_{ij}\bigg)
\frac{V_k\,\kappa\,\d \kappa}{R-\kappa\, \BR\cdot\BV}\bigg]\bigg|_{t'=t_\kappa}\,,
\end{align}
where $\BR$ and $\BV$ are to be evaluated at the corresponding retarded times.
The elastodynamic retarded potentials~(\ref{phi}) and (\ref{A}) fulfill
\begin{align}
\label{RP-rel}
A_{ijk}(\rr,t)=V_k(t_{\text{ret}})\,\phi_{ij}(\rr,t)\,.
\end{align}
The first $\delta_{ij}$-terms in Eqs.~(\ref{phi}) and (\ref{A}) have an analogous 
form as the 
original Li\'enard-Wiechert potentials of a point charge in the
electromagnetic theory\footnote{The original  Li\'enard-Wiechert potentials 
(scalar potential $\phi$ and vector potential $\BA$) of a point charge
read~\citep{Sommerfeld,Griffiths}:
\begin{align*}
\phi(\rr,t)=
\frac{q}{4\pi \epsilon_0}
\bigg[\frac{1}{R(t')-\BR(t')\cdot\BV(t')/c}\bigg]_{t'=t_c},\quad
\BA(\rr,t)=
\frac{q}{4\pi \epsilon_0 c^2 }
\bigg[\frac{\BV(t')}{R(t')-\BR(t')\cdot\BV(t')/c}\bigg]_{t'=t_c},
\end{align*}
where $q$ is the electric charge.
Here $t_c$ denotes the retarded time 
with respect to the velocity of light.
They fulfill the Lorentz gauge condition: $\dot{\phi}+c^2 \text{div}\BA=0$.
}
(see, e.g., \citep{LL,Jackson}). 
The Li\'enard-Wiechert tensor potential~(\ref{phi}) is analogous to the 
displacement vector, which is the Li\'enard-Wiechert vector potential, 
of a non-uniformly moving point force in elastodynamics found by~\citet{Lazar12}.
Due to the appearance of two velocities of the elastic waves, 
the elastodynamic 
Li\'enard-Wiechert tensor potentials have a more complicated
but rather straightforward structure.
Like the retarded elastic fields~(\ref{B-M-ret}) and (\ref{v-M-ret}), 
the Li\'enard-Wiechert tensor potentials~(\ref{phi}) and (\ref{A}) consist of three characteristic terms, again the first term
corresponds to the $S$-wave motion, the second term represents  the $P$-wave motion,
and the third term corresponds to a combination of $P$ and $S$ motion.
Thus, at the retarded positions the dislocation loop is the source of
$S$-, $P$- and the combination of $P$- and $S$-waves.

Finally, if we substitute Eqs.~(\ref{phi}) and (\ref{A}) in 
Eqs.~(\ref{B-M-L2}) and (\ref{v-M-L2}),
the elastic fields read in terms of the Li\'enard-Wiechert tensor potentials
\begin{align}
\label{B-M-L3}
\beta_{im}(\rr,t)&=
\frac{1}{4\pi\rho}\,C_{jkln}\, 
b_l\, \epsilon_{mnp}\,\pd_k
\bigg\{
\frac{1}{c^2_\TT}
\bigg[\oint_{L(t')}
\bigg(\delta_{ij}-\frac{R_iR_j}{R^2}\bigg)
 \frac{1}{R-\BR\cdot\BV/c_\TT}\, \d L_p(\Bs(t'))\bigg]\bigg|_{t'=t_\TT}
\nonumber\\
&\qquad\quad
+\frac{1}{c^2_\TL}
\bigg[
\oint_{L(t')}
\frac{R_iR_j}{R^2}\,
 \frac{1}{R-\BR\cdot\BV/c_\TL}\, \d L_p(\Bs(t'))\bigg]\bigg|_{t'=t_\TL}
\nonumber\\
&\qquad\quad
+\int_{1/c_\TL}^{1/c_\TT}
\bigg[\oint_{L(t')}
\bigg(\frac{3R_i R_j}{R^2}-\delta_{ij}\bigg)
\frac{\kappa}{R-\kappa\, \BR\cdot\BV}\, \d L_p(\Bs(t'))
\bigg]\bigg|_{t'=t_\kappa}\d \kappa
\bigg\}\nonumber\\
&\quad
+\frac{1}{4\pi}\, 
b_j\, \epsilon_{mkp}\,\pd_t
\bigg\{
\frac{1}{c^2_\TT}
\bigg[\oint_{L(t')}
\bigg(\delta_{ij}-\frac{R_iR_j}{R^2}\bigg)
 \frac{V_k}{R-\BR\cdot\BV/c_\TT}\, \d L_p(\Bs(t'))\bigg]\bigg|_{t'=t_\TT}
\nonumber\\
&\qquad\quad
+\frac{1}{c^2_\TL}
\bigg[\oint_{L(t')}
\frac{R_iR_j}{R^2}\,
 \frac{V_k}{R-\BR\cdot\BV/c_\TL}\, \d L_p(\Bs(t'))\bigg]\bigg|_{t'=t_\TL}
\nonumber\\
&\qquad\quad
+\int_{1/c_\TL}^{1/c_\TT}
\bigg[\oint_{L(t')}
\bigg(\frac{3R_i R_j}{R^2}-\delta_{ij}\bigg)
\frac{\kappa\, V_k}{R-\kappa\, \BR\cdot\BV}\, \d L_p(\Bs(t'))
\bigg]\bigg|_{t'=t_\kappa}\d \kappa
\bigg\}
\end{align}
and
\begin{align}
\label{v-M-L3}
v_{i}(\rr,t)&=\frac{1}{4\pi\rho}\, 
C_{jklm}\,
b_l\, \epsilon_{mnp}\,\pd_k
\bigg\{
\frac{1}{c^2_\TT}
\bigg[\oint_{L(t')}
\bigg(\delta_{ij}-\frac{R_iR_j}{R^2}\bigg)
 \frac{V_n}{R-\BR\cdot\BV/c_\TT}\, \d L_p(\Bs(t'))\bigg]\bigg|_{t'=t_\TT}
\nonumber\\
&\qquad\quad
+\frac{1}{c^2_\TL}
\bigg[\oint_{L(t')}
\frac{R_iR_j}{R^2}\,
 \frac{V_n}{R-\BR\cdot\BV/c_\TL}\, \d L_p(\Bs(t'))\bigg]\bigg|_{t'=t_\TL}
\nonumber\\
&\qquad\quad
+\int_{1/c_\TL}^{1/c_\TT}
\bigg[\oint_{L(t')}
\bigg(\frac{3R_i R_j}{R^2}-\delta_{ij}\bigg)
\frac{\kappa\, V_n}{R-\kappa\, \BR\cdot\BV}\, \d L_p(\Bs(t'))
\bigg]\bigg|_{t'=t_\kappa}\d \kappa
\bigg\}\,,
\end{align}
where again $\BR=\rr-\Bs(t')$ and $\BV=\dot{\Bs}(t')$ 
are to be evaluated at the corresponding retarded times
which must be determined by solving Eq.~(\ref{t-ret}).
The fields~(\ref{B-M-L3}) and (\ref{v-M-L3}) must be evaluated at 
some earlier times $t'$ (the retarded times) and for the corresponding
point $\Bs(t')$ on the dislocation loop.
Also the line element $\d L_p$ depends at every point $\Bs(t')$ on the
retarded times.
Thus, the elastic fields are given as line integrals around 
the dislocation line 
in terms of the Li\'enard-Wiechert tensor potentials of point sources.
Due to the retarded times in 
Eqs.~(\ref{B-M-L3}) and (\ref{v-M-L3}), $L(t')$ depends on
the retarded times and three curves $L(t_\TT)$, $L(t_\TL)$ and 
$L(t_\kappa)$ appear in Eqs.~(\ref{B-M-L3}) and (\ref{v-M-L3}).
Because of the explicit dependence of the retarded times on  $\Bs(t')$,
every point $\Bs(t')$ on the loop $L(t')$ depends on its own retarded time. 
For a dislocation loop, the time dependence of the elastic fields is 
based on a retardation due to the retarded times which are functions of
the variables $\rr$, $\Bs$, and $t$.
The elastic fields~(\ref{B-M-L3}) and (\ref{v-M-L3}) at the point $\rr$ and at
time $t$ receive contribution from
every point $\Bs(t_{\text{ret}})$ on the moving loop 
sending the elastic waves ($S$,$P$,mixed waves) at the retarded times 
($t_\TT,t_\TL,t_\kappa$).
For a moving loop these times are different not only due to different points
on the loop, but also because the loop moves.
The evaluation of the fields~(\ref{B-M-L3}) and (\ref{v-M-L3}) is far from 
a trivial task. Thus, Eqs.~(\ref{B-M-L3}) and (\ref{v-M-L3}) are 
complicated line integrals depending on the retarded times,
and time and spatial derivatives outside the line integrals.

Alternatively,
the substitution of Eqs.~(\ref{T-L}) and (\ref{I-L}) into 
the retarded elastic fields~(\ref{B-M-ret}) and (\ref{v-M-ret}) 
gives directly the fields~(\ref{B-M-L3}) and  (\ref{v-M-L3}). 
Thus, the tensor structure of ~(\ref{B-M-L3}) and  (\ref{v-M-L3})
is inherited from the tensor structure of 
the retarded elastic fields~(\ref{B-M-ret}) and (\ref{v-M-ret}).
Because the time-variable of the sources $\BT$ and $\BI$ 
in Eqs.~(\ref{B-M-ret}) and (\ref{v-M-ret}) is the retarded time $t_{\text{ret}}$, 
consequently the closed loop curve depends on the retarded times
$L(t_{\text{ret}})$ in Eqs.~(\ref{B-M-L3}) and  (\ref{v-M-L3}).  
Although the structure of Eqs.~(\ref{B-M-L3}) and  (\ref{v-M-L3}) is elegant from
the mathematical point of view, the evaluation of such expressions for a
non-uniformly moving dislocation loop is very complicated due to the 
dependence of the retarded times.
But this is the price we have to pay if we perform the integration in time 
and use the three-dimensional Green tensor~(\ref{GT}) 
in order to obtain the elastodynamic Li\'enard-Wiechert tensor potentials.
Since the relation of the retarded position to the present position 
of the dislocation loop is not in general, known, the 
elastic fields and the Li\'enard-Wiechert tensor potentials 
ordinarily permit only the evaluation of the fields in terms of the 
retarded positions and velocities of the dislocation loop.
The complexity of the fields~(\ref{B-M-L3}) and  (\ref{v-M-L3}) 
is hidden behind  the elegance of these formulae.

On the other hand,
if we substitute Eqs.~(\ref{T-L}) and (\ref{I-L}) into the Jefimenko type
formulae~(\ref{B-ret-Jef}) and (\ref{v-ret-Jef})
and carry out the integration in $\rr'$, we obtain
\begin{align}
\label{B-HF}
\beta_{im}(\rr,t)&=
-\frac{1}{4\pi\rho}\,C_{jkln}
b_l \epsilon_{mnp}
\bigg\{
\frac{1}{c^2_\TT}
\bigg[\oint_{L(t')}\!
\bigg(\frac{\delta_{ij} R_k +\delta_{jk} R_i +\delta_{ik} R_j}{R^2}
\nonumber\\
&\hspace{6cm}
-\frac{3R_iR_jR_k}{R^4}\bigg)
\frac{1}{R-\BR\cdot\BV/c_\TT}\, \d L_p(\Bs(t'))\bigg]\bigg|_{t'=t_\TT}\nonumber\\
&\qquad\quad
+\frac{1}{c^3_\TT}\pd_t 
\bigg[\oint_{L(t')}
\bigg(\delta_{ij}-\frac{R_iR_j}{R^2}\bigg)\frac{R_k}{R}
 \frac{1}{R-\BR\cdot\BV/c_\TT}\, \d L_p(\Bs(t'))\bigg]\bigg|_{t'=t_\TT}
\nonumber\\
&\qquad\quad
-\frac{1}{c^2_\TL}
\bigg[\oint_{L(t')}
\bigg(\frac{\delta_{jk} R_i +\delta_{ik} R_j}{R^2}
-\frac{3R_iR_jR_k}{R^4}\bigg) \frac{1}{R-\BR\cdot\BV/c_\TL}\, \d
L_p(\Bs(t'))\bigg]\bigg|_{t'=t_\TL}
\nonumber\\
&\qquad\quad
+\frac{1}{c^3_\TL}\pd_t
\bigg[\oint_{L(t')}
\frac{R_iR_jR_k }{R^3}\, \frac{1}{R-\BR\cdot\BV/c_\TL}\, \d L_p(\Bs(t'))\bigg]\bigg|_{t'=t_\TL}
\nonumber\\
&\ 
-\int_{1/c_\TL}^{1/c_\TT}
\bigg[\oint_{L(t')}
\bigg(\frac{\delta_{ij} R_k +3\delta_{jk} R_i +3\delta_{ik} R_j}{R^2}
-\frac{9R_iR_jR_k}{R^4}\bigg)
\frac{\kappa}{R-\kappa\, \BR\cdot\BV}\, \d L_p(\Bs(t'))
\bigg]\bigg|_{t'=t_\kappa}\d \kappa
\nonumber\\
&\qquad\quad
+\pd_t
\int_{1/c_\TL}^{1/c_\TT}
\bigg[\oint_{L(t')}
\bigg(\frac{3R_i R_j}{R^2}-\delta_{ij}\bigg)\frac{R_k}{R}
\frac{\kappa^2}{R-\kappa\, \BR\cdot\BV}\, \d L_p(\Bs(t'))
\bigg]\bigg|_{t'=t_\kappa}\d \kappa
\bigg\}\nonumber\\
&\quad
+\frac{1}{4\pi}\, 
b_j\, \epsilon_{mkp}\bigg\{
\frac{1}{c^2_\TT}\,\pd_t
\bigg[\oint_{L(t')}
\bigg(\delta_{ij}-\frac{R_iR_j}{R^2}\bigg)
 \frac{V_k}{R-\BR\cdot\BV/c_\TT}\, \d L_p(\Bs(t'))\bigg]\bigg|_{t'=t_\TT}
\nonumber\\
&\qquad\quad
+\frac{1}{c^2_\TL}\, \pd_t
\bigg[\oint_{L(t')}
\frac{R_iR_j}{R^2}\,
 \frac{V_k}{R-\BR\cdot\BV/c_\TL}\, \d L_p(\Bs(t'))\bigg]\bigg|_{t'=t_\TL}
\nonumber\\
&\qquad
+ \pd_t\int_{1/c_\TL}^{1/c_\TT}
\bigg[\oint_{L(t')}
\bigg(\frac{3R_i R_j}{R^2}-\delta_{ij}\bigg)
\frac{\kappa\, V_k}{R-\kappa\, \BR\cdot\BV}\, \d L_p(\Bs(t'))
\bigg]\bigg|_{t'=t_\kappa}\d \kappa\bigg\}
\end{align}
and
\begin{align}
\label{v-HF}
v_{i}(\rr,t)&=
-\frac{1}{4\pi\rho}\,C_{jklm}
b_l \epsilon_{mnp}
\bigg\{
\frac{1}{c^2_\TT}
\bigg[\oint_{L(t')}\!
\bigg(\frac{\delta_{ij} R_k +\delta_{jk} R_i +\delta_{ik} R_j}{R^2}
\nonumber\\
&\hspace{6cm}
-\frac{3R_iR_jR_k}{R^4}\bigg)
\frac{V_n}{R-\BR\cdot\BV/c_\TT}\, \d L_p(\Bs(t'))\bigg]\bigg|_{t'=t_\TT}\nonumber\\
&\qquad\quad
+\frac{1}{c^3_\TT}\pd_t 
\bigg[\oint_{L(t')}
\bigg(\delta_{ij}-\frac{R_iR_j}{R^2}\bigg)\frac{R_k}{R}
 \frac{V_n}{R-\BR\cdot\BV/c_\TT}\, \d L_p(\Bs(t'))\bigg]\bigg|_{t'=t_\TT}
\nonumber\\
&\qquad\quad
-\frac{1}{c^2_\TL}
\bigg[\oint_{L(t')}
\bigg(\frac{\delta_{jk} R_i +\delta_{ik} R_j}{R^2}
-\frac{3R_iR_jR_k}{R^4}\bigg) \frac{V_n}{R-\BR\cdot\BV/c_\TL}\, \d
L_p(\Bs(t'))\bigg]\bigg|_{t'=t_\TL}
\nonumber\\
&\qquad\quad
+\frac{1}{c^3_\TL}\pd_t
\bigg[\oint_{L(t')}
\frac{R_iR_jR_k }{R^3}\,
 \frac{V_n}{R-\BR\cdot\BV/c_\TL}\, \d L_p(\Bs(t'))\bigg]\bigg|_{t'=t_\TL}
\nonumber\\
&\ 
-\int_{1/c_\TL}^{1/c_\TT}
\bigg[\oint_{L(t')}
\bigg(\frac{\delta_{ij} R_k +3\delta_{jk} R_i +3\delta_{ik} R_j}{R^2}
-\frac{9R_iR_jR_k}{R^4}\bigg)
\frac{\kappa\, V_n }{R-\kappa\, \BR\cdot\BV}\, \d L_p(\Bs(t'))
\bigg]\bigg|_{t'=t_\kappa}\d \kappa
\nonumber\\
&\qquad\quad
+\pd_t 
\int_{1/c_\TL}^{1/c_\TT}
\bigg[\oint_{L(t')}
\bigg(\frac{3R_i R_j}{R^2}-\delta_{ij}\bigg)\frac{R_k}{R}
\frac{\kappa^2\, V_n }{R-\kappa\, \BR\cdot\BV}\, \d L_p(\Bs(t'))
\bigg]\bigg|_{t'=t_\kappa}\d \kappa
\bigg\}\,.
\end{align}
Eqs.~(\ref{B-HF}) and (\ref{v-HF}) are analogous to the Heaviside-Feynman
formulae~\citep{Heaviside,Feynman} of electromagnetic theory and we may call
them the Heaviside-Feynman type formulae for a dislocation loop. 
They are more complicated than the original Heaviside-Feynman 
formulae~\citep{Heaviside,Feynman} (see also~\citep{Schott,Jackson,HM,Eyges}) 
for a point charge in electrodynamics\footnote{In electrodynamics, 
the Heaviside-Feynman formulae for the electric field strength $\BE$ and 
the magnetic field strength $\BB$ 
of a non-uniformly moving point charge 
are of the form~\citep{HM,Jackson}:
\begin{align*}
\BE(\rr,t)&=
\frac{q}{4\pi \epsilon_0}
\bigg(
\bigg[\frac{\RR}{R^2 (R- \BR\cdot\BV/c)}\bigg]_{t_c}
+\frac{1}{c}\,\pd_t 
\bigg[\frac{\RR}{R (R- \BR\cdot\BV/c)}\bigg]_{t_c}
-\frac{1}{c^2}\,\pd_t 
\bigg[\frac{\BV}{R- \BR\cdot\BV/c}\bigg]_{t_c}
\bigg)\,,\\
\BB(\rr,t)&=
\frac{q}{4\pi \epsilon_0 c^2}
\bigg(
\bigg[\frac{\BV\times \RR}{R^2 (R- \BR\cdot\BV/c)}\bigg]_{t_c}
+\frac{1}{c}\,\pd_t 
\bigg[\frac{\BV\times\RR}{R (R- \BR\cdot\BV/c)}\bigg]_{t_c}
\bigg)\,.
\end{align*}
}.
Eqs.~(\ref{B-HF}) and (\ref{v-HF})
contain terms without derivatives and terms with 
time-derivatives. 
They still possess a quite complicated tensor structure and line integrals
depending on the retarded times.
The causal dependencies of the elastic fields of a
non-uniformly moving dislocation loop are described by
retarded line integrals along the loop.

\section{Static limit of a dislocation loop}
In this section, we give the static limit of the Li\'enard-Wiechert
tensor potentials and of the elastic distortion of a dislocation loop 
as a check of the above results.

In order to carry out the static limit, we set
$\BV=0$ and $\Bs(t')=\rr'$ and 
we substitute Eq.~(\ref{c}) and $\lambda=2\mu\nu/(1-2\nu)$
into Eq.~(\ref{phi}). If we perform the integration in $\kappa$ and 
arrange in proper order the appearing terms, we find 
\begin{align}
\phi_{ij}=G_{ij}
\end{align}
with 
\begin{align}
\label{G-s}
G_{ij}=\frac{1}{16\pi\mu(1-\nu)}\,\big[2(1-\nu)\delta_{ij}\Delta
-\pd_i\pd_j\big] R\, ,
\end{align}
where $\RR=\rr-\rr'$.
Eq.~(\ref{G-s}) is the static three-dimensional Green tensor of the Navier 
equation~\citep{Teodosiu,Lardner}.
That means that the Li\'enard-Wiechert
tensor potential $\phi_{ij}$ reduces to the static elastic Green tensor 
in the static limit.
In this manner, we find from Eq.~(\ref{B-M-L2}) the so-called
Mura formula~(e.g. \citep{Mura,deWit2,Li}) as static limit
\begin{align}
\label{B-M-s}
\beta_{im}(\rr)=
\oint_L\epsilon_{mnp} \,C_{jkln}\, G_{ij,k}\, b_l\, \d L'_p\, ,
\end{align}
which gives the elastic distortion of a static dislocation loop.

On the other hand, the static limit of Eq.~(\ref{B-HF}) is obtained as
\begin{align}
\label{B-HF-stat}
\beta_{im}(\rr)
=-\frac{b_l}{8\pi(1-\nu)}\oint_L\epsilon_{mnp}
\bigg\{(1-2\nu)\frac{\delta_{il}R_n+\delta_{in}R_l-\delta_{ln}R_i}{R^3}
+\frac{3R_iR_lR_n}{R^5}\bigg\} \, \d L_p'\,,
\end{align}
which is the explicit form of Eq.~(\ref{B-M-s}).

\section{Straight dislocations}
We derive now the two-dimensional Li\'enard-Wiechert tensor potentials 
and the elastic fields of straight dislocations in our general framework.
We choose the dislocation line $\ell_z$ parallel to the $z$-axis.
The dislocation density and dislocation current tensors of 
edge dislocations are given by
\begin{align}
\label{dd-e}
T_{ijk}=b_i\, \epsilon_{jkz}\,\ell_z\, \delta(\BR)\, ,\qquad
I_{ij}=b_i\, \epsilon_{jkz} V_k\, \ell_z\, \delta(\BR)\, .
\end{align}
For a screw dislocation, 
the dislocation density and dislocation current tensors read
\begin{align}
\label{dd-s}
T_{zjk}=b_z\, \epsilon_{jkz}\,\ell_z\, \delta(\BR)\, ,\qquad
I_{zj}=b_z\, \epsilon_{jkz} V_k \,\ell_z\, \delta(\BR)\, ,
\end{align}
where $\BR=\rr-\Bs(t) \in \R^2$ and $i,j,k=x,y$.
The tensors~(\ref{dd-e}) and (\ref{dd-s}) fulfill, respectively, the relations
\begin{align}
\label{rel-dd}
I_{ij}=V_k\, T_{ijk}\,,\qquad
I_{zj}=V_k\, T_{zjk}\,.
\end{align}
Eq.~(\ref{rel-dd}) gives a relation between the dislocation current tensor, 
the dislocation density tensor and the dislocation velocity vector,
which is valid for a single straight dislocation. 
\citet{Guenther67} and \citet{Teodosiu70} 
derived such a relation between $\BI$ and $\BT$ for 
uniformly moving dislocations, that means for constant dislocation velocity $\BV$.
If the relation~(\ref{rel-dd}) is valid, then $\BI$ is a convection
dislocation current.

If the material is infinitely extended, the two-dimensional elastodynamic 
Green tensor of plane-strain reads~\citep{Eringen75,Kausel}
\begin{align}
\label{GT-2D}
G_{ik}(\rr,t)&=\frac{1}{2\pi\rho }\, 
\Bigg\{
\frac{x_i x_k}{r^4}\,
\bigg(
\frac{\big[2t^2-r^2/c^2_\TL\big]}{\sqrt{t^2-r^2/c^2_\TL}} 
\, H\big(t-r/c_\TL\big)
-\frac{\big[2t^2-r^2/c^2_\TT\big]}{\sqrt{t^2-r^2/c^2_\TT}}
\ H\big(t-r/c_\TT\big) 
\bigg)\nonumber\\
&\qquad\qquad
-\frac{\delta_{ik}}{r^2}\, 
\bigg(
\sqrt{t^2-r^2/c^2_\TL}\, H\big(t-r/c_\TL\big)
-\frac{t^2}{\sqrt{t^2-r^2/c^2_\TT}}
\, H\big(t-r/c_\TT\big)
\bigg)\Bigg\}
\end{align}
and the elastodynamic Green tensor of anti-plane strain is given by
\begin{align}
\label{GT-zz}
G_{zz}(\rr,t)=\frac{1}{2\pi\rho c_\TT^2}\, 
\frac{H\big(t-r/c_\TT\big)}{\sqrt{t^2-r^2/c^2_\TT}}\, ,
\end{align}
where $H(.)$ denotes the Heaviside step function and $r=\sqrt{x^2+y^2}$.

If we substitute Eqs.~(\ref{GT-2D}) and (\ref{dd-e}) in Eqs.~(\ref{B-M}) and 
(\ref{v-M}) and perform the integration in $\rr'$, we find for the elastic fields
(no summation over $z$ in Eqs.~(\ref{B-S}) and (\ref{v-S}))
\begin{align}
\beta_{im}(\rr,t)&=
\label{B-S}
\epsilon_{mnz}\, C_{jkln}\, \pd_k \phi_{ij}\, 
b_l\,  \ell_z
+
\rho\, \pd_t A_{ijk}\, b_j \epsilon_{mkz} \,  \ell_z\,
\end{align}
and
\begin{align}
v_i(\rr,t)=
\label{v-S}
C_{jklm}\, \pd_k A_{ijn}\, 
 b_l \epsilon_{mnz}\,  \ell_z\,.
\end{align}
Using the definitions of the elastodynamic 
Li\'enard-Wiechert tensor potentials~(\ref{phi-0}) and (\ref{A-0}) and the 
Green tensor~(\ref{GT-2D}), we have here 
introduced the two-dimensional Li\'enard-Wiechert tensor potentials
\begin{align}
\label{phi-S}
\phi_{ij}(\rr,t)&=\frac{1}{2\pi\rho}
\bigg[
\int_{-\infty}^{t_{{\TL}}}  \bigg(\frac{ R_iR_j}{R^4}\, \frac{\bar{t}^2}{S_\TL}
+\bigg(\frac{ R_iR_j-\delta_{ij}\, R^2}{R^4}\bigg)S_\TL\bigg)\d t'\nonumber\\
&\qquad\quad
-\int_{-\infty}^{t_{{\TT}}}  \bigg(\frac{ R_iR_j}{R^4}\, S_\TT
+\bigg(\frac{ R_iR_j-\delta_{ij}\, R^2}{R^4}\bigg)
\frac{\bar{t}^2}{S_\TT}\bigg)\d t'
\bigg]\, ,\\
\label{A-S}
A_{ijk}(\rr,t)&=\frac{1}{2\pi\rho}
\bigg[
\int_{-\infty}^{t_{{\TL}}}  V_k(t')
\bigg(\frac{ R_iR_j}{R^4}\, \frac{\bar{t}^2}{S_\TL}
+\bigg(\frac{ R_iR_j-\delta_{ij}\, R^2}{R^4}\bigg)S_\TL\bigg)
\d t'\nonumber\\
&\qquad\quad
-\int_{-\infty}^{t_{{\TT}}} V_k(t') \bigg(\frac{ R_iR_j}{R^4}\, S_\TT
+\bigg(\frac{ R_iR_j-\delta_{ij}\, R^2}{R^4}\bigg)
\frac{\bar{t}^2}{S_\TT}\bigg)\d t'
\bigg]\, .
\end{align}
The notation here is 
\begin{align}
\label{Not-eg}
&R_i=x_i-s_i(t')\, ,\qquad \bar{t}=t-t'\, , \qquad
&S^2_{\TT}=\bar{t}^2-\frac{{R}^2}{c^2_{\TT}}\, ,
\qquad
S^2_{\TL}=\bar{t}^2-\frac{{R}^2}{c^2_{\TL}}\, .
\end{align}
The two retarded times $t_\TT=t'$ and $t_\TL=t'$
are the roots of $S_\TT^2=0$ and $S_\TL^2=0$, respectively,
which are less than $t$.
The solving of the conditions $S_\TT^2=0$ and $S_\TL^2=0$
is for a general motion non-trivial and can be very complicated.
For subsonic dislocations the solutions for the retarded times
$t_\TT$ and $t_\TL$ are unique.
The general expressions~(\ref{B-S})--(\ref{A-S}) contain the 
fields for gliding as well as climbing dislocations. 
If $\BV\| \Bb$, the expressions~(\ref{B-S})--(\ref{A-S}) describe a 
gliding edge dislocation 
and if $\BV\perp \Bb$, Eqs.~(\ref{B-S})--(\ref{A-S}) give the fields 
of a climbing edge dislocation (see, e.g., \citep{Lazar2011}). 
For these cases, we can recover from Eqs.~(\ref{B-S})--(\ref{A-S})
the explicit expressions given by~\citet{Mura64,Lardner} and \citet{Lazar2011}.

For a screw dislocation, if we
insert Eqs.~(\ref{GT-zz}) and (\ref{dd-s}) in Eqs.~(\ref{B-M}) and 
(\ref{v-M}), and performing the integration in $\rr'$, we obtain for the elastic fields
\begin{align}
\beta_{zm}(\rr,t)&=
\label{B-scr}
\epsilon_{mnz}\, C_{zkzn}\, \pd_k \phi_{zz}\, 
b_z\,  \ell_z
+
\rho\, \pd_t A_{zzk}\, b_z \epsilon_{mkz} \,  \ell_z\,
\end{align}
and
\begin{align}
v_z(\rr,t)=
\label{v-scr}
C_{zkzm}\, \pd_k A_{zzn}\, 
 b_z \epsilon_{mnz}\,  \ell_z\,.
\end{align}
Using the definitions Eqs.~(\ref{phi}) and (\ref{A}) and the Green 
function~(\ref{GT-zz}), the two-dimensional Li\'enard-Wiechert potentials of anti-plane strain are
\begin{align}
\phi_{zz}(\rr,t)&=\frac{1}{2\pi\rho c_\TT^2}
\int_{-\infty}^{t_{{\TT}}}
\frac{1}{S_\TT}\,\d t'\, ,\qquad
\label{A-scr}
A_{zzk}(\rr,t)=\frac{1}{2\pi\rho c_\TT^2}
\int_{-\infty}^{t_{{\TT}}}
\frac{V_k(t')}{S_\TT}\,\d t'\, .
\end{align}
It should be noted that the index $z$ in Eqs.~(\ref{B-scr})--(\ref{A-scr})
is a fixed index (no summation over $z$)  
due to the reduction from 3D to the anti-plane strain problem. 
From the general expressions~(\ref{B-scr})--(\ref{A-scr}), we 
can reduce all the field components given by \citet{Eshelby53,Nabarro,Lardner}, and
\citet{Lazar2011}.

The two-dimensional Li\'enard-Wiechert potentials~(\ref{phi-S}), (\ref{A-S}),
and (\ref{A-scr}) are time-integrals over the history of the motion and thus,
they possess an afterglow. 
For that reason, a straight dislocation is haunted by its past as 
\citet{Eshelby51} mentioned.
Nevertheless, for the two-dimensional 
Li\'enard-Wiechert potentials~(\ref{phi-S}), (\ref{A-S})
and (\ref{A-scr}) 
we are left to evaluate time-integrals of considerable complexity which only 
in some simple cases yield results of elementary functions in a closed form.
Also the calculation of the retarded times is not a trivial task.
For straight dislocations, the static limit of the elastic fields 
of non-uniformly moving straight dislocations was given in~\citep{Lazar2011}.

\section{Near-field approximation of straight dislocations}

In the near-field approximation, it is important to determine the character
of the singularities of
Eqs.~(\ref{phi-S}), (\ref{A-S}) and (\ref{A-scr})
at $\RR\rightarrow 0$. 
Thus, we have to calculate only the near-field approximation 
of the integrals in Eqs.~(\ref{phi-S}), (\ref{A-S}) and (\ref{A-scr}).
We follow the near-field approximation for cylindrical waves given 
by~\citet{Hilbert,Whitham} and \citet{Barton}.
We consider here dislocations which are at rest and
at $t'=0$ they start to move.
This problem can be viewed as the superposition
of the static equilibrium problem for $t<0$ with 
the dynamic problem for $t>0$.
Since the wave propagation nature of the solution is of interest, 
we consider the superposition-related problem 
obtained by subtracting the equilibrium solution from the complete solution.
Therefore, 
the lower integration limits of the Li\'enard-Wiechert potentials~(\ref{phi-S}), 
(\ref{A-S}) and (\ref{A-scr}) are changed from $-\infty$ to 0 
(see also \citep{Brock82,Brock86,XM82,XM84}).

We start with the near-field approximation for a screw dislocation.
In the near-field approximation, the retardation is negligible.
Thus, we have $\BV(t')\approx \BV(t)$ and $\RR=\rr-\Bs(t')\approx\rr-\Bs(t)$.
The two-dimensional Li\'enard-Wiechert potentials~(\ref{A-scr}) 
reduce to
\begin{align}
\label{phi-appr}
2\pi\rho c_\TT^2\, \phi_{zz}(\rr,t)
&\approx
\int_{0}^{t}\frac{H(\tau-R/c_\TT)}{\sqrt{\tau^2-R^2/c^2_\TT}}\,\d t'\,,
\quad \tau=t-t' \nonumber\\
&=
\int_{0}^{t}
\frac{H(\tau-R/c_\TT)}{\sqrt{\tau^2-R^2/c^2_\TT}}\,\d \tau\,\nonumber\\
&=H(t-R/c_\TT)
\int_{R/c_\TT}^{t}
\frac{\d \tau}{\sqrt{\tau^2-R^2/c^2_\TT}}\,\nonumber\\
&=H(t-R/c_\TT)\, 
\ln\bigg[\frac{c_\TT t}{R}+\sqrt{\frac{c^2_\TT t^2}{R^2}-1}\bigg]\nonumber\\ 
&\approx H(t-R/c_\TT)\, \ln \frac{2 c_\TT t}{R}+\OO(R^2)
\end{align}
and 
$A_{zzk}(\rr,t) =V_k(t) \phi_{zz}(\rr,t)$.
For $t\gg R/c_\TT$, Eq.~(\ref{phi-appr}) simplifies to (see also~\citep{Barton})
\begin{align}
\phi_{zz}(\rr,t)&\approx\frac{1}{2\pi\rho c_\TT^2}
\,\ln \frac{2 c_\TT t}{R}\, ,\qquad
\label{A-scr-near}
A_{zzk}(\rr,t)\approx\frac{1}{2\pi\rho c_\TT^2}
\, V_k(t)\ln \frac{2 c_\TT t}{R}\, .
\end{align}
If we substitute (\ref{A-scr-near}) into Eqs.~(\ref{B-scr}) and (\ref{v-scr}),
we obtain the singular terms of the elastic fields of a moving screw dislocation
\begin{align}
\label{Bz-near}
\beta_{zm}(\rr,t)=-\epsilon_{mnz}\,\frac{b_z \ell_z}{2\pi\rho c_\TT^2}
\bigg(C_{zkzn} \frac{R_k}{R^2}-\rho\, V_n(t)\, \frac{\RR\cdot\BV(t)}{R^2}
-\rho\, \dot{V}_n(t)\ln  \frac{2 c_\TT t}{R}\bigg)
\end{align}
and
\begin{align}
\label{vz-near}
v_{z}(\rr,t)=-\epsilon_{mnz}\,\frac{b_z \ell_z}{2\pi\mu}
\, C_{zkzm} \frac{R_k V_n(t)}{R^2}\,  .
\end{align}
Explicitly, they read
\begin{align}
\label{Bzx-near}
\beta_{zx}(\rr,t)&=-\frac{b_z \ell_z}{2\pi}
\bigg\{\bigg(1-\frac{V^2_y(t)}{c_\TT^2}\bigg)\frac{R_y}{R^2}
-\frac{V_x(t)V_y(t)}{c_\TT^2}\, \frac{R_x}{R^2}
-\frac{\dot{V}_y(t)}{c^2_\TT}\ln  \frac{2 c_\TT t}{R}\bigg\}\\
\label{Bzy-near}
\beta_{zy}(\rr,t)&=\frac{b_z \ell_z}{2\pi}
\bigg\{\bigg(1-\frac{V^2_x(t)}{c_\TT^2}\bigg)\frac{R_x}{R^2}
-\frac{V_x(t)V_y(t)}{c_\TT^2}\, \frac{R_y}{R^2}
-\frac{\dot{V}_x(t)}{c^2_\TT}\ln  \frac{2 c_\TT t}{R}\bigg\}\\
\label{vz-near2}
v_{z}(\rr,t)&=\frac{b_z \ell_z}{2\pi}
\,\frac{R_y V_x(t)-R_x V_y(t)}{R^2}\,.
\end{align}
It is obvious in Eqs.~(\ref{Bzx-near})--(\ref{vz-near2})
that a screw dislocation contains a $1/R$ singularity and a logarithmic 
singularity in the near-field. The acceleration terms give the 
logarithmic singularity.

Also Eqs.~(\ref{Bzx-near}) and (\ref{Bzy-near}) 
contain the correct static limit with $\BV=0$ and $\Bs(t)=\text{constant}$ 
for the elastic distortion of a screw dislocation with the Burgers vector $\Bb=(0,0,b_z)$
given by~\citet{deWit73}
\begin{align}
\label{Bzx}
\beta_{zx}&=-\frac{b_z}{2\pi}\, \frac{y}{r^2}\,,\\
\label{Bzy}
\beta_{zy}&=\frac{b_z}{2\pi}\, \frac{x}{r^2}\,. 
\end{align}
For convenience we have chosen $\Bs=0$.

For the near-fields of edge dislocations, we need the following integral 
approximations
\begin{align}
\label{int-edge1}
\int_{0}^{t}
\frac{H(\tau-R/c)\, \tau^2}{\sqrt{\tau^2-R^2/c^2}}\,\d t'
&=
 \int_{0}^{t}
\frac{H(\tau-R/c)\, \tau^2}{\sqrt{\tau^2-R^2/c^2}}\,\d \tau\,\nonumber\\
&=H(t-R/c)
\int_{R/c}^{t}
\frac{\tau^2\, \d \tau}{\sqrt{\tau^2-R^2/c^2}}\,\nonumber\\
&=H(t-R/c)\bigg(
\frac{R^2}{2 c^2}\,
\ln\bigg[\frac{c t}{R}+\sqrt{\frac{c^2 t^2}{R^2}-1}\bigg]
+\frac{t}{2 }\sqrt{t^2-\frac{R^2}{c^2}}\bigg)\nonumber\\
&\approx H(t-R/c)\bigg(\frac{R^2}{2 c^2}\,\ln \frac{2 c t}{R} 
+\frac{t^2}{2} -\frac{R^2}{4 c^2} 
+\OO(R^4)\bigg)
\end{align}
and
\begin{align}
\label{int-edge2}
\int_{0}^{t}H(\tau-R/c)\, \sqrt{\tau^2-R^2/c^2}\,\d t'
&=
 \int_{0}^{t}
H(\tau-R/c)\, \sqrt{\tau^2-R^2/c^2}\,\d \tau\,\nonumber\\
&=H(t-R/c)
\int_{R/c}^{t}
\sqrt{\tau^2-R^2/c^2}\, \d \tau\nonumber\\
&=H(t-R/c)\bigg(
-\frac{R^2}{2 c^2}\,
\ln\bigg[\frac{ct}{R}+\sqrt{\frac{c^2 t^2}{R^2}-1}\bigg]
+\frac{t}{2 }\sqrt{t^2-\frac{R^2}{c^2}}\bigg)\nonumber\\
&\approx H(t-R/c)\bigg(-\frac{R^2}{2 c^2}\,\ln \frac{2 c t}{R} 
+\frac{t^2}{2}  -\frac{R^2}{4 c^2} 
+\OO(R^4)\bigg).
\end{align}
Using Eqs.~(\ref{int-edge1}) and (\ref{int-edge2}), we obtain for the
Li\'enard-Wiechert tensor potentials~(\ref{phi-S}) and (\ref{A-S})
\begin{align}
\label{phi-e-app1}
2\pi\rho\, \phi_{ij}(\rr,t)
&\approx 
H(t-R/c_\TL)\Bigg\{
\frac{R_iR_j}{R^4} t \sqrt{t^2-\frac{R^2}{c^2_\TL}}\nonumber\\
&\hspace{2cm}
+\delta_{ij}\Bigg(
\frac{1}{2c^2_\TL}\ln\bigg[\frac{c_\TL t}{R}+\sqrt{\frac{c^2_\TL t^2}{R^2}-1}\bigg]
-\frac{t}{2 R^2 }\sqrt{t^2-\frac{R^2}{c^2_\TL}}\Bigg)\Bigg\}\nonumber\\
&\, -H(t-R/c_\TT)\Bigg\{
\frac{R_iR_j}{R^4} t\sqrt{t^2-\frac{R^2}{c^2_\TT}}\nonumber\\
&\hspace{2cm}
-\delta_{ij}\Bigg(
\frac{1}{2c^2_\TT}\ln\bigg[\frac{c_\TT t}{R}+\sqrt{\frac{c^2_\TT t^2}{R^2}-1}\bigg]
+\frac{t}{2 R^2 }\sqrt{t^2-\frac{R^2}{c^2_\TT}}\Bigg)\Bigg\}\
\end{align}
and $A_{ijk}(\rr,t)=V_k(t)\, \phi_{ij}(\rr,t)$. 
If $t\gg R/c_\TT$ and $t\gg R/c_\TL$, then (\ref{phi-e-app1}) simplifies to
\begin{align}
\label{phi-e-app2}
\phi_{ij}(\rr,t)\approx
\frac{1}{4\pi\rho}\bigg[
\bigg(\frac{1}{c^2_\TT}-\frac{1}{c^2_\TL}\bigg)\frac{R_iR_j}{ R^2}+
\delta_{ij}\bigg(\frac{1}{c^2_\TT}\,\ln \frac{2 c_\TT t}{R}
+\frac{1}{c^2_\TL}\,\ln \frac{2 c_\TL t}{R}\bigg)\bigg]
\end{align}
and $A_{ijk}(\rr,t)=V_k(t)\, \phi_{ij}(\rr,t)$.
We calculate
\begin{align}
\label{grad-phi}
\pd_k \phi_{ij}(\rr,t)=
-\frac{1}{8\pi\mu(1-\nu)}\bigg[(3-4\nu)\, \delta_{ij}\, \frac{R_k}{R^2}
-\delta_{ik}\frac{R_j}{R^2}-\delta_{jk}\frac{R_i}{R^2}
+2\,\frac{R_i R_j R_k}{R^4}\bigg]\, ,
\end{align}
$\pd_k A_{ijk}(\rr,t)=V_k(t)\pd_k \phi_{ij}(\rr,t)$ and
\begin{align}
\label{dot-A}
\pd_t A_{ijk}(\rr,t)&=
\frac{\dot{V}_k(t)}{8\pi\mu(1-\nu)}\, \frac{R_iR_j}{R^2}
+\frac{\dot{V}_k(t)\,\delta_{ij}}{4\pi\rho}\bigg[
\frac{1}{c^2_\TT}\,\ln \frac{2 c_\TT t}{R}
+\frac{1}{c^2_\TL}\,\ln \frac{2 c_\TL t}{R}\bigg]\nonumber\\
&\quad
+\frac{V_k(t)}{8\pi\mu(1-\nu)}\bigg[
(3-4\nu)\, \delta_{ij}\, \frac{\BR\!\cdot\! \BV}{R^2}
-\frac{R_i V_j+R_j V_i}{R^2}
+2\,\frac{R_i R_j 
\BR\!\cdot\! \BV}{R^4}\bigg]\, .
\end{align}
The first term in Eq.~(\ref{dot-A}) is non-singular and, thus, it 
does not contribute to the singular near-field.
Substituting~(\ref{grad-phi}) and (\ref{dot-A}) into 
Eqs.~(\ref{B-S}) and (\ref{v-S}), we find for the singular near-fields of the elastic
fields of edge dislocations
\begin{align}
\label{beta-near}
\beta_{im}(\rr,t)&=
-\frac{\epsilon_{mnz} b_l \ell_z }{8\pi\mu(1-\nu)}
\bigg(C_{jkln}
\bigg[(3-4\nu)\, \delta_{ij}\, \frac{R_k}{R^2}
-\delta_{ik}\frac{R_j}{R^2}-\delta_{jk}\frac{R_i}{R^2}
+2\,\frac{R_i R_j R_k}{R^4}\bigg]
\nonumber\\
&\quad
-\rho V_n(t)
\bigg[
(3-4\nu)\, \delta_{il}\, \frac{\BR\!\cdot\! \BV}{R^2}
-\frac{R_i V_l+R_l V_i}{R^2}
+2\,\frac{R_i R_l \BR\!\cdot\! \BV}{R^4}\bigg]\bigg)
\nonumber\\
&\quad
+\frac{\epsilon_{mnz} b_i \ell_z\, \dot{V}_n(t)}{4\pi}
\bigg[
\frac{1}{c^2_\TT}\,\ln \frac{2 c_\TT t}{R}
+\frac{1}{c^2_\TL}\,\ln \frac{2 c_\TL t}{R}\bigg]
\end{align}
and
\begin{align}
\label{v-near}
v_{i}(\rr,t)&=
-\frac{\epsilon_{mnz} b_l \ell_z C_{jklm} V_n(t) }{8\pi\mu(1-\nu)}
\bigg((3-4\nu)\, \delta_{ij}\, \frac{R_k}{R^2}
-\delta_{ik}\frac{R_j}{R^2}-\delta_{jk}\frac{R_i}{R^2}
+2\,\frac{R_i R_j R_k}{R^4}\bigg)\,,
\end{align}
where $\RR=\rr-\Bs(t)$. 
It can be seen in Eqs.~(\ref{beta-near}) and (\ref{v-near})
that edge dislocations contain a $1/R$ singularity and a logarithmic 
singularity in the near-field approximation. 
The acceleration terms in Eq.~(\ref{beta-near}) give the 
logarithmic singularity. 
Eqs.~(\ref{beta-near}) and (\ref{v-near}) are valid for 
climbing and gliding edge dislocations.

It is also important to note that Eq.~(\ref{beta-near}) 
gives the correct static limit with $\BV=0$ and $\Bs(t)=\text{constant}$
for the 
elastic distortion of an edge dislocation with Burgers vector $\Bb=(b_x,0,0)$ 
given by~\citet{deWit73}
\begin{align}
\label{Bxx}
\beta_{xx}&=-\frac{b_x}{4\pi(1-\nu)}\, \frac{y}{r^2}
\bigg[(1-2\nu)+2\, \frac{x^2}{r^2}\bigg]\,,\\
\label{Byx}
\beta_{yx}&=-\frac{b_x}{4\pi(1-\nu)}\, \frac{x}{r^2}
\left[(1-2\nu)+2\, \frac{y^2}{r^2}\right]\,,\\
\label{Bxy}
\beta_{xy}&=\frac{b_x}{4\pi(1-\nu)}\, \frac{x}{r^2}
\left[(3-2\nu)-2\,\frac{y^2}{r^2}\right]\,,\\
\label{Byy}
\beta_{yy}&=-\frac{b_x}{4\pi(1-\nu)}\, \frac{y}{r^2}
\left[(1-2\nu)-2\, \frac{x^2}{r^2}\right]\,.
\end{align}
Again we have chosen $\Bs=0$.

It is important to mention
that the history of the motion is not visible in the singularities of the
non-uniformly moving dislocations~(\ref{Bz-near}), (\ref{vz-near}),
(\ref{beta-near}), and (\ref{v-near}).
The singular terms are determined by the current situation.
Both screw and edge dislocations are characterized by two types of 
singularities, namely, $1/R$ and logarithmic singularities.
The logarithmic singularity is characteristic for the acceleration motion.
The logarithmic singularities (acceleration terms) appear only
in the elastic distortions~(\ref{Bz-near}) and (\ref{beta-near}) 
and not in the elastic velocity terms~(\ref{vz-near}) and (\ref{v-near}).
The character of the singularities of the near fields is in 
agreement with the results given by~\citet{XM85,XM1990} and  \citet{XM2003,XM2008}. 
A near-field approximation of singular integrals giving 
simple coefficients of the singular terms has been used.
In this simple near-field approximation, we have neglected the retardation.

\section{
The Mura dislocation tensor potentials,
the retarded dislocation tensor potentials
and the Li\'enard-Wiechert tensor potentials of a dislocation loop}

On the other hand, \citet{Mura64b,Mura} has written the elastic fields in terms
of so-called dislocation tensor potentials in the following way
\begin{align}
\beta_{im}(\rr,t)&=
\label{B-Mura}
\epsilon_{mnp}\big( C_{jkln}\, \pd_k \phi_{ijlp}
+\rho\, \pd_t A_{ijnjp}\big)\,,\\
v_i(\rr,t)&=
\label{v-Mura}
\epsilon_{mnp} C_{jklm}\, \pd_k A_{ijnlp}\, .
\end{align}
The dislocation tensor potentials $\phi_{ijkl}$ and $A_{ijkmn}$, 
introduced by~\citet{Mura64b,Mura}, are defined by (see also~\citep{Bonilla})
\begin{align}
\phi_{ijkl}(\rr,t)&=
\label{phi-Mura}
\int_{-\infty}^t\int_{-\infty}^\infty G_{ij}(\rr-\rr', t-t')\, 
\alpha_{kl}(\rr',t')\, \d \rr'\, \d t'\,,\\
A_{ijkmn}(\rr,t)&=
\label{A-Mura}
\int_{-\infty}^t\int_{-\infty}^\infty G_{ij}(\rr-\rr', t-t')\, 
V_{kmn}(\rr',t')\, \d \rr'\, \d t'\,,
\end{align}
and they fulfill the following wave equations
\begin{align}
\label{phi-pde-Mura}
&\big[\delta_{ik}\,\rho \, \pd_{tt}-C_{ijkl}\pd_j\pd_l\big] \phi_{kmpq}=
\delta_{im}\,\alpha_{pq}\, ,\\
\label{A-pde-Mura}
&\big[\delta_{ik}\,\rho \, \pd_{tt}-C_{ijkl}\pd_j\pd_l\big] A_{kmnpq}=
\delta_{im}\, V_{npq}\, ,
\end{align}
where $V_{kmn}$ is Mura's dislocation velocity tensor~\citep{Mura63,Mura63b}. 
Although $V_{kmn}$ possesses 27 components, only 9 components which are
skew-symmetric in some indices
($V_{kmn}=-V_{nmk}$ and $A_{ijkmn}=-A_{ijnmk}$)
enter Eqs.~(\ref{B-Mura}) and (\ref{v-Mura}).
For a straight dislocation with
Burgers vector $b_m$, dislocation line direction $\ell_n$ and dislocation velocity
$V_k$, it reads: $V_{kmn}= V_k b_m \ell_n\, \delta(\RR)=V_k\alpha_{mn}$.
Here $\alpha_{ij}$ denotes the usual dislocation density tensor~(\ref{alpha}).
The dislocation current tensor~(\ref{I}) can be written in terms of 
Mura's dislocation velocity tensor as follows
\begin{align}
\label{IV-rel}
I_{ij}=\epsilon_{jkl} V_{kil} \,.
\end{align}
Substituting Eqs.~(\ref{alpha}) and (\ref{IV-rel}) into Eq.~(\ref{BI2}),
the Bianchi identity~(\ref{BI2}) reads
\begin{align}
\label{BI3}
\dot{\alpha}_{il}+V_{kil,k}-V_{lik,k}=0\, ,
\end{align}
which leads to the following 'gauge' or side condition for the dislocation
tensor potentials (see also~\citep{Mura})
\begin{align}
\label{GC}
\dot{\phi}_{ijkl}+A_{ijmkl,m}-A_{ijlkm,m}=0\, .
\end{align}

If we substitute the Green tensor~(\ref{GT}) into Eqs.~(\ref{phi-Mura}) and 
(\ref{A-Mura}) and perform the integration in time $t'$, we find
the retarded dislocation tensor potentials\footnote{
In electrodynamics, the retarded electromagnetic potentials were originally 
introduced by~\citet{Lorenz} and they read~\citep{Griffiths,Jackson}:
\begin{align*}
\phi(\rr,t)=
\frac{1}{4\pi \epsilon_0}
\int_\VV \frac{\rho(\rr',t-R/c)}{R}\, \d \rr',\quad
\BA(\rr,t)=
\frac{1}{4\pi \epsilon_0 c^2 }
\int_\VV \frac{\BJ(\rr',t-R/c)}{R}\, \d \rr'\,,
\end{align*}
where $\rho$ is the electric charge density and $\BJ$ 
denotes the electric current density vector.
The idea of a retarded scalar potential was first developed by \citet{Lorenz61} 
in 1861 when studying waves in the theory of elasticity.
The retarded potentials fulfill the 
Lorentz gauge condition: $\dot{\phi}+c^2 \text{div}\BA=0$ 
(see, e.g.,~\citep{Griffiths,Jefimenko}).}
\begin{align}
\label{phi-RP}
\phi_{ijkl}(\rr,t)&=
\frac{1}{4\pi\rho}\, \int_\VV
\bigg\{
\frac{1}{c^2_\TT}\bigg(\frac{\delta_{ij}}{R}
-\frac{R_i R_j}{R^3}\bigg) \alpha_{kl}(\rr',t_\TT)
+\frac{1}{c^2_\TL}\,\frac{R_i R_j}{R^3}\,\alpha_{kl}(\rr',t_\TL)
\nonumber\\
&\hspace{17mm}
+\bigg(\frac{3R_i R_j}{R^3}-\frac{\delta_{ij}}{R}\bigg)
\int_{1/c_\TL}^{1/c_\TT} \kappa\,\alpha_{kl}(\rr',t_\kappa)\, \d \kappa\bigg\}\,\d \rr'
\end{align}
and
\begin{align}
\label{A-RP}
A_{ijkmn}(\rr,t)&=
\frac{1}{4\pi\rho }\, \int_\VV
\bigg\{
\frac{1}{c^2_\TT}\bigg(\frac{\delta_{ij}}{R}
-\frac{R_i R_j}{R^3}\bigg) V_{kmn}(\rr',t_\TT)
+\frac{1}{c^2_\TL}\,\frac{R_i R_j}{R^3}\,V_{kmn}(\rr',t_\TL)
\nonumber\\
&\hspace{17mm}
+\bigg(\frac{3R_i R_j}{R^3}-\frac{\delta_{ij}}{R}\bigg)
\int_{1/c_\TL}^{1/c_\TT} \kappa\,V_{kmn}(\rr',t_\kappa)\, \d \kappa\bigg\}\,
\d \rr'\,,
\end{align}
where the retarded times $t_\TT$, $t_\TL$ and $t_\kappa$ 
are given by Eqs.~(\ref{tT})--(\ref{tkappa}).
Because the integrands are evaluated at the retarded times, 
these fields are called retarded dislocation tensor potentials.
They are the causal solutions of the inhomogeneous 
Navier equations~(\ref{phi-pde-Mura})
and (\ref{A-pde-Mura}).
The retarded dislocation tensor 
potentials~(\ref{phi-RP}) and (\ref{A-RP}) satisfy the gauge condition~(\ref{GC}),
this can be easily checked.
In the static case, 
the retarded dislocation tensor potentials~(\ref{phi-RP})
and (\ref{A-RP}) reduce to
\begin{align}
\label{RP-stat}
\phi_{ijkl}(\rr)
=\frac{1}{16 \pi\mu (1-\nu)}\int_\VV
\frac{1}{R}\bigg[(3-4\nu)\delta_{ij}
+\frac{R_iR_j}{R^2}\bigg]\alpha_{kl}(\rr') \, \d \rr'\,,
\end{align}
and $A_{ijkmn}(\rr)=0$.
On the other hand, if we substitute Eqs.~(\ref{phi-RP}) and (\ref{A-RP}) 
into (\ref{B-Mura}) and (\ref{v-Mura}) and use the relations~(\ref{IV-rel}),
(\ref{alpha-inv}), and (\ref{v-M-ret}),
we obtain the Jefimenko type equations~(\ref{B-ret-Jef}) and (\ref{v-ret-Jef}).

If we compare Eqs.~(\ref{B-Mura}) and (\ref{v-Mura}) with 
(\ref{B-M-L2}) and (\ref{v-M-L2}), we find 
\begin{align}
\label{LW-loop}
\phi_{ijmn}(\rr,t)=\oint_{L(t')}\phi_{ij}\, b_m \, \d L_n(\Bs(t'))\,,\qquad
A_{ijkmn}(\rr,t)=\oint_{L(t')}A_{ijk }\, b_m \, \d L_n(\Bs(t'))\,,
\end{align}
that means that the dislocation tensor potentials are given as 
line integrals of the three-dimensional elastodynamic 
Li\'enard-Wiechert tensor potentials of a so-called
`point dislocation source'~(\ref{phi}) and (\ref{A}) along 
the dislocation line $L(t')$.
The dislocation tensor potentials of a dislocation loop~(\ref{LW-loop})
may be identified with the actual Li\'enard-Wiechert tensor potentials of
a dislocation loop.
If we substitute Eqs.~(\ref{phi}) and (\ref{A}) into 
(\ref{LW-loop}), they read explicitly
\begin{align}
\label{phi-loop}
\phi_{ijmn}(\rr,t)&=
\frac{b_m}{4\pi\rho}\,
\bigg\{
\frac{1}{c^2_\TT}
\bigg[
\oint_{L(t')}
\bigg(\delta_{ij}-\frac{R_iR_j}{R^2}\bigg)
 \frac{1}{R-\BR\cdot\BV/c_\TT}\, \d L_n(\Bs(t'))\bigg]\bigg|_{t'=t_\TT}
\nonumber\\
&\qquad\quad
+\frac{1}{c^2_\TL}
\bigg[\oint_{L(t')}
\frac{R_iR_j}{R^2}\,
 \frac{1}{R-\BR\cdot\BV/c_\TL}\, \d L_n(\Bs(t'))\bigg]\bigg|_{t'=t_\TL}
\nonumber\\
&\qquad\quad
+\int_{1/c_\TL}^{1/c_\TT}
\bigg[\oint_{L(t')}
\bigg(\frac{3R_i R_j}{R^2}-\delta_{ij}\bigg)
\frac{\kappa}{R-\kappa\, \BR\cdot\BV}\, \d L_n(\Bs(t'))
\bigg]\bigg|_{t'=t_\kappa}\d \kappa\bigg\}
\end{align}
and
\begin{align}
\label{A-loop}
A_{ijkmn}(\rr,t)&=
\frac{b_m}{4\pi\rho}\, 
\bigg\{
\frac{1}{c^2_\TT}\bigg[
\oint_{L(t')}
\bigg(\delta_{ij}-\frac{R_iR_j}{R^2}\bigg)
 \frac{V_k}{R-\BR\cdot\BV/c_\TT}\, \d L_n(\Bs(t'))\bigg]\bigg|_{t'=t_\TT}
\nonumber\\
&\qquad\quad
+\frac{1}{c^2_\TL}
\bigg[\oint_{L(t')}
\frac{R_iR_j}{R^2}\,
 \frac{V_k}{R-\BR\cdot\BV/c_\TL}\, \d L_n(\Bs(t'))\bigg]\bigg|_{t'=t_\TL}
\nonumber\\
&\qquad\quad
+\int_{1/c_\TL}^{1/c_\TT}
\bigg[\oint_{L(t')}
\bigg(\frac{3R_i R_j}{R^2}-\delta_{ij}\bigg)
\frac{\kappa\, V_k}{R-\kappa\, \BR\cdot\BV}\, \d L_n(\Bs(t'))
\bigg]\bigg|_{t'=t_\kappa}\d \kappa\bigg\}\,,
\end{align}
where $\BV=\BV(t')$, $\BR=\BR(t')$, $\Bs(t')$ and $L(t')$ have to be 
computed at the corresponding retarded times.
Thus, the Li\'enard-Wiechert tensor potentials of a dislocation
loop~(\ref{phi-loop}) and (\ref{A-loop}) are line integrals 
of the  Li\'enard-Wiechert tensor potentials of 
a dislocation point source~(\ref{phi}) and (\ref{A}) 
integrated over three loop curves $L(t_\TT)$,  $L(t_\TL)$,
and  $L(t_\kappa)$ at the corresponding retarded times and with the integrands
$\d L_n(\Bs(t_\TT))$, $\d L_n(\Bs(t_\TL))$, and $\d L_n(\Bs(t_\kappa))$.
Eqs.~(\ref{phi-loop}) and (\ref{A-loop}) are retarded line integrals.
Thus, the Li\'enard-Wiechert tensor potentials of a dislocation loop
are the line integrals of a `dislocation point source' acting on $\Bs(t')$ 
integrated over the dislocation loop $L(t')$ and evaluated at the
corresponding retarded times.
The elastic fields of a non-uniformly moving 
dislocation loop are obtained by substituting
Eqs.~(\ref{phi-loop}) and (\ref{A-loop}) into~(\ref{B-Mura}) and (\ref{v-Mura})
reproducing the formulae~(\ref{B-M-L3}) and (\ref{v-M-L3}).
The fields produced by a dislocation loop 
can therefore be computed by first determining 
the Li\'enard-Wiechert tensor potentials~(\ref{phi-loop}) and (\ref{A-loop}) 
and then obtaining the elastic fields by means of 
the relations~(\ref{B-Mura}) and (\ref{v-Mura}).
More directly, the  Li\'enard-Wiechert tensor potentials of a dislocation
loop~(\ref{phi-loop}) and (\ref{A-loop}) can be obtained by substituting
the dislocation density tensor and Mura's dislocation velocity tensor 
of a dislocation loop
\begin{align}
\label{A-L}
\alpha_{ij}(\rr,t)&=b_i \oint_{L(t)}\delta(\rr-\Bs(t))\, \d
L_j(\Bs(t))\,,\\
\label{V-L}
V_{kij}(\rr,t)&=b_i \oint_{L(t)}V_k(t)\, \delta(\rr-\Bs(t))\, \d
L_j(\Bs(t))\,,
\end{align}
into Eqs.~(\ref{phi-Mura}) and (\ref{A-Mura}) or 
in Eqs.~(\ref{phi-RP}) and (\ref{A-RP}).
Eqs.~(\ref{A-L}) and (\ref{V-L}) can be read-off from Eqs.~(\ref{T-L}) and 
(\ref{I-L}), using the relations~(\ref{alpha-inv}) and (\ref{IV-rel}).

In the static case,
the Li\'enard-Wiechert tensor potentials of a dislocation loop~(\ref{phi-loop})
and (\ref{A-loop}) reduce to
\begin{align}
\label{LW-stat}
\phi_{ijmn}(\rr)
=\frac{b_m}{16 \pi\mu (1-\nu)}\oint_L\frac{1}{R}
\bigg[(3-4\nu)\delta_{ij}
+\frac{R_iR_j}{R^2}\bigg] \, \d L_n'\,,
\end{align}
and $A_{ijkmn}(\rr)=0$, where $\BR=\rr-\rr'$.

Comparing Eqs.~(\ref{B-Mura}) and (\ref{v-Mura}) with (\ref{B-S}) and (\ref{v-S}), 
we obtain 
for the two-dimensional Li\'enard-Wiechert tensor potentials of plane strain
\begin{align}
\phi_{ijmz}=\phi_{ij} b_m \ell_z\,,\qquad
A_{ijkmz}=A_{ijk} b_m \ell_z\, ,
\end{align}
and with  (\ref{B-scr}) and (\ref{v-scr}), 
we obtain 
for the two-dimensional Li\'enard-Wiechert tensor potentials of anti-plane strain
\begin{align}
\phi_{zzzz}=\phi_{zz} b_z \ell_z\,,\qquad
A_{zzkzz}=A_{zzk} b_z \ell_z\,.
\end{align}
From the condition (\ref{GC}), we get 
\begin{align}
\label{GC-2}
\dot{\phi}_{ij}+A_{ijk,k}=0
\end{align}
and
\begin{align}
\label{GC-3}
\dot{\phi}_{zz}+A_{zzk,k}=0\,.
\end{align}
A `gauge' condition like~(\ref{GC-3}) was already mentioned by~\citet{Lardner}.

Therefore, it is obvious that the elastodynamical 
Li\'enard-Wiechert tensor potentials can be obtained from 
Mura's dislocation tensor potentials. Especially, for a dislocation loop
the Mura dislocation tensor potentials give directly the 
Li\'enard-Wiechert tensor potentials as retarded line integrals along  loop 
curves computed at the corresponding retarded times.

\section{Conclusions}
In this work, we have investigated the fundamentals of 
the non-uniform motion of dislocations based on electromagnetic analogies.
We have examined and solved to following items:
\begin{itemize}
\item 
retarded elastic fields/Jefimenko type equations for dislocation fields
\item
retarded dislocation tensor potentials
\item
3-D Li\'enard-Wiechert tensor potentials of point dislocation sources 
\item
3-D Li\'enard-Wiechert tensor potentials of a dislocation loop 
\item
Heaviside-Feynman type equations for a dislocation loop
\item
2-D Li\'enard-Wiechert tensor potentials of straight dislocations
\item
singularities of the near-fields of straight dislocations.
\end{itemize}
This analysis provides the treatment of general dislocation motion
(with inertia effects) in terms of retarded fields.
We have introduced the elastodynamic 
Li\'enard-Wiechert tensor potentials as fundamental 
quantities for the non-uniform motion of dislocations.
For the general motion of a dislocation loop, the solution is obtained in terms 
of Li\'enard-Wiechert tensor potentials depending on the retarded times  
and in a more closed form as line integrals (Heaviside-Feynman type formulae). 
The Jefimenko type and Heaviside-Feynman type equations contain only time
derivatives and no spatial derivatives.
Thus, for the general motion of a loop the solution is evaluated 
in a closed form.
It is concluded that causal dependencies of the elastic fields and
Li\'enard-Wiechert tensor potentials of a non-uniformly moving dislocation loop
are described by retarded line integrals along the dislocation loop.
The non-uniform motion of dislocations leads to retardation effects
which are important in the interaction between dislocations. 
Thus, the interaction between dislocations during the dislocation motion 
is not instantaneous, without being limited by the speed of the sound waves.
We think that these results will open a new window for a better understanding
of the non-uniform motion of dislocation fields.
Especially, they can be used in simulations of discrete dislocation dynamics.
Retardation effects should be considered in dislocation dynamics 
simulations at high strain rates and they become important at extremely high frequencies.

\section*{Acknowledgement}
The author gratefully acknowledges the grants obtained from the 
Deutsche Forschungsgemeinschaft (Grant Nos. La1974/2-1, La1974/3-1). 
The author wishes to express his gratitude to 
Prof. H.O.K. Kirchner for stimulating discussions, criticism and useful remarks
on an earlier version of the paper.

\end{document}